\documentclass[submission, Phys]{SciPost}
\usepackage{subcaption}
\usepackage[utf8]{inputenc} % input encodings (allow UTF-8 input)
\usepackage[T1]{fontenc} 	% selecting font encodings (use 8-bit T1 fonts)
\usepackage[english]{babel} % multilingual support (English language/hyphenation)

\usepackage[bitstream-charter]{mathdesign}
\usepackage{geometry} 		% flexible and complete interface to document dimensions
\usepackage{amsmath} 		% math package (American Mathematical Society)
\usepackage{mathtools} 		% math package (fixes various deficiencies of amsmath)
\usepackage{float} 			% floating objects such as figures and tables
\usepackage{graphicx} 		% enhanced support for graphics
\usepackage{tabularx} 		% tabulars with adjustable-width columns
\usepackage{booktabs} 		% professional-quality tables
\usepackage{color, xcolor} 	% foreground and background colour management
\usepackage{pdfpages} 		% inclusion of external multi-page PDF documents
\usepackage{extarrows} 		% extra arrows beyond those provided in amsmath
\usepackage{multirow} 		% create tabular cells spanning multiple rows
\usepackage{multicol} 		% intermix single and multiple columns
\usepackage{enumitem} 		% control layout of itemize, enumerate, description
\usepackage{xspace} 		% define commands that appear not to eat spaces
\usepackage{stackrel} 		% enhancement to the \stackrel command
\usepackage{tikz} 			% create PostScript and PDF graphics
\usepackage{braket} 		% Dirac bra-ket notation
\usepackage{bm} 			% access bold symbols in maths mode
\usepackage{tensor} 		% typeset tensors (tensor-style super- and subscripts)
\usepackage{slashed} 		% slash through characters (Feynman slash notation)
\usepackage{siunitx} 		% SI units package (typesetting values with units)
\usepackage{lastpage} 		% reference last page
\usepackage{cite} 			% improved citation handling
\usepackage[normalem]{ulem} % package for underlining
\usepackage{fontawesome} 	% access to web-related icons
\usepackage{tocloft} 		% control over the typography of the TOC, LOF, and LOT
\usepackage{titlesec} 		% interface to sectioning commands (various title styles)
\usepackage{doi} 			% create correct hyperlinks for DOI numbers
\usepackage{hyperref} 		% hypertext links (handel cross-referencing commands)
\usepackage[most]{tcolorbox} 					% coloured and framed text boxes
\usepackage[nameinlink, capitalize]{cleveref} 	% intelligent cross-referencing
\usepackage[nottoc, notlot, notlof]{tocbibind} 	% add references/index/contents to TOC
\usepackage[ruled, vlined]{algorithm2e} 		% floating algorithm environment
\usepackage{makecell}
\usepackage[makeroom]{cancel}
\usepackage{feynmf}

\makeatletter
\def\BState{\State\hskip-\ALG@thistlm}
\makeatother

\makeatletter
\@ifundefined{pdfoutput}{}{\DeclareGraphicsRule{*}{mps}{*}{}}
\makeatother

\makeatletter
\DeclareRobustCommand*{\bfseries}{%
   \not@math@alphabet\bfseries\mathbf
   \fontseries\bfdefault\selectfont
   \boldmath
}
\makeatother

\hypersetup{
	pdftitle={CaloDREAM},
	pdfauthor={Favaro L., Ore A., Palacios-Schweitzer S., Plehn T.},
	colorlinks=true, 			% false: boxed links, true: colored links
	linkcolor={red!50!black}, 	% color of internal links (sections, pages, etc.)
	citecolor={blue!50!black}, 	% color of citation links (links to bibliography)
	urlcolor={blue!80!black} 	% color of URL links (external links)
} 
\DeclareSymbolFont{usualmathcal}{OMS}{cmsy}{m}{n}
\DeclareSymbolFontAlphabet{\mathcal}{usualmathcal}

\SetArgSty{textnormal}
\SetKwComment{Comment}{{\small\#}~}{}
\SetCommentSty{mycommfont}

\setitemize{itemsep=0pt, parsep=0pt} 				% adjust itemize environment
\setenumerate{itemsep=0pt, parsep=0pt} 				% adjust enumerate environment
\setitemize{itemsep=2pt,topsep=2pt,parsep=0pt,partopsep=0pt,leftmargin=*}
\setenumerate{itemsep=0pt,topsep=2pt,parsep=0pt,partopsep=0pt,labelindent=3pt,leftmargin=*}
\usepackage{amsmath}
 
\usepackage{amsthm} 		% math package (typesetting theorems using AMS style)
\theoremstyle{definition}

\usetikzlibrary{arrows, shapes.geometric, arrows.meta, shapes, decorations.pathreplacing, fit, patterns, patterns.meta,positioning,calc}
\usepackage{tikz} 			% create PostScript and PDF graphics

\definecolor{Rcolor}{HTML}{E99595}
\definecolor{Gcolor}{HTML}{C5E0B4}
\definecolor{Gcolor_light}{HTML}{F1F8ED}
\definecolor{Bcolor}{HTML}{9DC3E6}
\definecolor{Ycolor}{HTML}{FFE699}
\definecolor{Ycolor_light}{HTML}{FFF7DE}
\newcommand{\tikznode}[2]{%
\ifmmode%

\tikz[remember picture,baseline=(#1.base),inner sep=0pt] \node (#1) {$#2$};%
\else
\tikz[remember picture,baseline=(#1.base),inner sep=0pt] \node (#1) {#2};%
\fi}

\tikzstyle{expr} = [rectangle, rounded corners=0.3ex, minimum width=1.5cm, minimum height=1cm, text centered, align=center, inner sep=0, fill=Ycolor, font=\large, draw]

\tikzstyle{small_cinn} = [double arrow, double arrow head extend=0cm, double arrow tip angle=130, inner sep=0, align=center, minimum width=1.1cm, minimum height=0.5cm, fill=Rcolor, draw]

\tikzstyle{small_cinn_black} = [small_cinn, minimum height=1.5cm, fill=black]

\tikzstyle{transformer} = [rectangle, rounded corners, minimum width=6cm, minimum height=2.4cm, font=\large, fill=Gcolor_light, draw]

\tikzstyle{attention} = [rectangle, rounded corners=0.3ex, minimum width=5.5cm, minimum height=1.2cm, align=center, fill=Gcolor, draw, font=\large]

\tikzstyle{transformer_huge} = [rectangle, rounded corners, minimum width=8.5cm, minimum height=2.4cm, font=\large, fill=Gcolor_light, draw]

\tikzstyle{attention_huge} = [rectangle, rounded corners=0.3ex, minimum width=8cm, minimum height=1.2cm, align=center, fill=Gcolor, draw, font=\large]

\tikzstyle{txt_huge} = [align=center, font=\Huge, scale=2]
\tikzstyle{txt} = [align=center, font=\LARGE, minimum height=1cm]

\tikzstyle{arrow} = [thick,-{Latex[scale=1.0]}, line width=0.2mm, color=black]
\tikzstyle{line} = [thick, line width=0.2mm, color=black]
\tikzstyle{bkg} = [minimum width=22cm, minimum height=7.5cm, font=\large, fill=white]

\tikzstyle{encoder_black} = [trapezium, fill=black, rotate=270, minimum height=4cm, minimum width=3.2cm,align=center,draw]
\tikzstyle{encoder} = [trapezium, fill=Bcolor, rotate=270, minimum height=4cm, minimum width=3cm,align=center,draw]
\tikzset{trapezium stretches=true}
\tikzstyle{cinn} = [double arrow, double arrow head extend=0cm, double arrow tip angle=130, shape border rotate=90, inner sep=0, align=center, minimum width=2.1cm, minimum height=2.3cm, fill=Rcolor, draw,font=\LARGE]
\tikzstyle{cinn_black} = [cinn, minimum height=2.5cm, fill=black]
\tikzstyle{crc} = [circle, rounded corners=0.3ex, minimum width=1.5cm, minimum height=1cm, text centered, align=center, inner sep=0, fill=white, font=\LARGE, draw] 
\definecolor{red_cb}{HTML}{e41a1c}
\definecolor{blue_cb}{HTML}{377eb8}
\definecolor{green_cb}{HTML}{4daf4a}
\definecolor{purple_cb}{HTML}{984ea3}
\definecolor{orange_cb}{HTML}{ff7f00}

\definecolor{EmeraldGreen}{HTML}{1ea78d}
\definecolor{EnglishRed}{HTML}{b02427}
\newcommand{\einc}{E_\text{inc}}

\newcommand{\pd}{p_\text{data}}
\newcommand{\Langle}{\bigl\langle}
\newcommand{\Rangle}{\bigr\rangle}
\newcommand{\XLangle}{\Bigl\langle}
\newcommand{\XRangle}{\Bigr\rangle}
\newcommand{\XXLangle}{\biggl\langle}
\newcommand{\XXRangle}{\biggr\rangle}

\newcommand{\qqquad}{\qquad\quad}

\newcommand\one{\leavevmode\hbox{\small1\normalsize\kern-.33em1}}
\newcommand{\R}{\mathbb{R}} 				% real numbers
\newcommand{\normal}{\mathcal{N}}

\newcommand{\uniform}{\mathcal{U}}

\newcommand{\loss}{\mathcal{L}} 	% loss value

\newcommand{\geant}{\textsc{Geant4}\xspace}

\newcommand{\arXiv}[2][]{%
	\ifthenelse{\equal{#1}{}}%
	{\href{http://arxiv.org/abs/#2}{arXiv:#2}}%
	{\href{http://arxiv.org/abs/#2}{arXiv:#2~[#1]}}}

\newcommand{\mev}{\text{MeV}}

\def\slashchar#1{\setbox0=\hbox{$#1$}           % set a box for #1
   \dimen0=\wd0                                 % and get its size
   \setbox1=\hbox{/} \dimen1=\wd1               % get size of /
   \ifdim\dimen0>\dimen1                        % #1 is bigger
      \rlap{\hbox to \dimen0{\hfil/\hfil}}      % so center / in box
      #1                                        % and print #1
   \else                                        % / is bigger
      \rlap{\hbox to \dimen1{\hfil$#1$\hfil}}   % so center #1
      /                                         % and print /
   \fi}

\def\mathswitchr#1{\relax\ifmmode{\mathrm{#1}}\else$\mathrm{#1}$\xspace\fi}
\def\mathswitch#1{\relax\ifmmode#1\else$#1$\xspace\fi}

\DeclarePairedDelimiterX{\infdivx}[2]{[}{]}{%
  #1\;\delimsize\|\;#2%
}

\graphicspath{{./figs/}}
\begin{document}

\begin{center}{\Large \textbf{
CaloDREAM --- \\ Detector Response Emulation via Attentive flow Matching
}}\end{center}

\begin{center}
  Luigi Favaro\textsuperscript{1,2}, 
  Ayodele Ore\textsuperscript{1}, 
  Sofia Palacios Schweitzer\textsuperscript{1}, 
  and Tilman Plehn\textsuperscript{1}
\end{center}

\begin{center}
{\bf 1} Institut f\"ur Theoretische Physik, Universit\"at Heidelberg, Germany\\
{\bf 2} CP3, Universit\'e catholique de Louvain, Louvain-la-Neuve, Belgium\\
\end{center}

\begin{center}
\today
\end{center}

% For convenience during refereeing: line numbers
%\linenumbers

\section*{Abstract}
{\bf Detector simulations are an exciting application
of modern generative networks.
Their sparse high-dimensional data combined with the 
required precision poses a serious challenge. 
We show how combining Conditional Flow Matching
with transformer elements allows us to simulate the 
detector phase space reliably. Namely, we use an autoregressive transformer to simulate the energy of each layer, and a vision transformer for the high-dimensional voxel distributions. 
We show how dimension reduction via 
latent diffusion allows us to train 
more efficiently and how diffusion networks can be evaluated 
faster with bespoke solvers. 
We showcase our framework, CaloDREAM, on datasets 2 and 3 of the CaloChallenge.}

\vspace{-1pt}
\noindent\rule{\textwidth}{1pt}
\tableofcontents\thispagestyle{fancy}
\noindent\rule{\textwidth}{1pt}
%\vspace{10pt}

\clearpage
%%%%%%%%%%%%%%%%%%%%%%%%%%%%%%%%%%%%%%%%%%%%%%%%%%%
\section{Introduction}
\label{sec:intro}

Simulations are the way we compare theory predictions to LHC data,
allowing us to draw conclusions about fundamental theory from complex
scattering data~\cite{Campbell:2022qmc,Butter:2022rso}. The modular
simulation chain starts from the hard interaction and progresses
through particle decays, QCD radiation, hadronization, hadron decays,
to the interaction of all particles with the detector. Currently, the
last step is turning into a bottleneck in speed and
precision. Generating calorimeter showers with
\geant~\cite{Agostinelli:2002hh,1610988,ALLISON2016186}, based on
first principles, requires a large fraction of the computing budget.
Without significant progress, this simulation step will be the
limiting factor for all analyses at the HL-LHC~\cite{HEPSoftwareFoundation:2018fmg, CERN-LHCC-2022-005}.

Modern machine learning is transforming the way we simulate LHC
data~\cite{Plehn:2022ftl}.  In the past few years
we have seen successful applications to all steps in the simulation
chain~\cite{Butter:2022rso}, phase space
integration~\cite{Bendavid:2017zhk,Klimek:2018mza,Chen:2020nfb,Gao:2020vdv,Bothmann:2020ywa,Gao:2020zvv,Danziger:2021eeg,Heimel:2022wyj,Janssen:2023ahv,Bothmann:2023siu,Heimel:2023ngj};
parton
showers~\cite{deOliveira:2017pjk,Andreassen:2018apy,Bothmann:2018trh,Dohi:2020eda,Buhmann:2023pmh,Leigh:2023toe,Mikuni:2023dvk,Buhmann:2023zgc};
hadronization~\cite{Ilten:2022jfm,Ghosh:2022zdz,Chan:2023ume,Bierlich:2023zzd};
detector
simulations~\cite{Paganini:2017hrr,deOliveira:2017rwa,Paganini:2017dwg,Erdmann:2018kuh,Erdmann:2018jxd,Belayneh:2019vyx,Buhmann:2020pmy,ATL-SOFT-PUB-2020-006,Buhmann:2021lxj,Krause:2021ilc,ATLAS:2021pzo,Krause:2021wez,Buhmann:2021caf,Chen:2021gdz,Adelmann:2022ozp,Mikuni:2022xry,ATLAS:2022jhk,Krause:2022jna,Cresswell:2022tof,Diefenbacher:2023vsw,Hashemi:2023ruu,Xu:2023xdc,Diefenbacher:2023prl,Buhmann:2023bwk,Buckley:2023rez,Mikuni:2023tqg,Amram:2023onf,Diefenbacher:2023flw,FaucciGiannelli:2023fow,Pang:2023wfx,Leigh:2023zle,Buhmann:2023kdg,Birk:2023efj,Schnake:2024mip},
and end-to-end event
generation~\cite{Otten:2019hhl,Hashemi:2019fkn,DiSipio:2019imz,Butter:2019cae,Alanazi:2020klf,Butter:2021csz,Butter:2023fov, Butter:2023ira},
including inverse
simulations~\cite{Datta:2018mwd,Bellagente:2019uyp,Andreassen:2019cjw,Bellagente:2020piv,Backes:2022vmn,Leigh:2022lpn,Raine:2023fko,Shmakov:2023kjj,Ackerschott:2023nax,Diefenbacher:2023wec,Huetsch:2024quz}
and simulation-based
inference~\cite{Bieringer:2020tnw,Butter:2022vkj,Heimel:2023mvw}.
While these new concepts and tools have the potential to transform LHC 
simulations, we need to ensure that these networks and their technical strengths can be understood. This is the only way that we can 
systematically improve the LHC simulation chain~\cite{Diefenbacher:2020rna,Butter:2021csz,Winterhalder:2021ave,Nachman:2023clf,Das:2023ktd}, 
without endangering the key role it plays in, essentially, every LHC analysis.
% Most notably, we have to ensure that LHC simulations remain first-principle and are not replaced by data-driven modelling.
Most notably, we must avoid the case where effects of interest are absorbed into LHC simulations as a result of data-driven modelling.
This means that for now 
we always assume that networks used in LHC simulations are trained 
on simulations and controlled by comparing to simulations. 

In this paper, we will apply cutting-edge generative networks to
calorimeter shower simulations. The high-dimensional phase spaces of
calorimeter showers are a challenge to the established normalizing
flows or INNs~\cite{Ernst:2023qvn}, and different variants of diffusion networks
appear to be the better-suited
architecture~\cite{Amram:2023onf}.
This is in spite of the fact that diffusion networks are, typically,
slower in the forward generation and do not allow for an efficient
likelihood extraction. In addition to showing that these
networks are able to simulate sparse phase space 
signals like calorimeter showers, we will explore which 
phase space dimensionalities we can describe with 
full-dimensional latent spaces and how a dimension-reduced
latent representation affects the network performance.

Given the \geant benchmark presented in 
Sec.~\ref{sec:setup}, we will see 
that a factorized approach is most promising. In Sec.~\ref{sec:dream} we 
first introduce a Conditional Flow Matching (CFM) network combined 
with an autoregressive transformer to learn the layer energies. Next, we combine it with a 
3-dimensional vision transformer to learn the shower shapes.
This combination can be trained on 
datasets~2 and~3 of the CaloChallenge to 
generate high-fidelity calorimeter showers.
In this application the step from dataset~2
to dataset~3 motivates a switch from full-dimensional voxel
representations to a dimension-reduced latent
space~\cite{Cresswell:2022tof}. In Sec.~\ref{sec:res} we study, in some detail, 
how the full-dimension generative network encodes the 
calorimeter shower information for both datasets.
To alleviate the computational challenges, we also show how a lower-dimensional 
latent representation helps us describe high-dimensional data like
the Calo Challenge dataset~3 and how the CFM networks can sample 
more efficiently. 
For quantitative benchmarking of the learned 
phase space distribution we employ a learned classifier test, indicating 
that the network precision for both datasets is at the per-cent level and 
the loss in precision from a reduced latent space is controlled, including 
its only failure mode, which are the sparsity distributions.

%%%%%%%%%%%%%%%%%%%%%%%%%%%%%%%%%%%%%%%%%%%%%%%%%%%
\section{Data and preprocessing}
\label{sec:setup}

To benchmark our new network architectures, we use dataset~2
(DS2)~\cite{CaloChallenge_ds2} and dataset~3
(DS3)~\cite{CaloChallenge_ds3} of the CaloChallenge
2022~\cite{Krause:2024avx}.  Each set consists of 200k
\geant~\cite{Agostinelli:2002hh} electron showers: 100k for training/validation and 100k for testing. Showers are simulated over a
log-uniform incident energy range
\begin{align}
    \label{eq:einc-range}
    \einc = 10^3~...~10^6~\mev \; .
\end{align}
The physical detector has a cylindrical geometry with alternating
layers of absorber and active material, altogether 90 layers. The
voxelization following Ref.~\cite{calochallenge} combines an active
layer and an absorber layer resulting in 45 concentric cylindrical
layers.

The particle originating the shower always enters at the (0,0,0)
location and defines the $z$-axis of the coordinate system. The number
of readout cells per layer is defined in a polar coordinate system and it is different for DS2 and DS3. DS2 has a total of 6480 voxels: 144
voxels per layer, each divided into 16 angular and 9 radial bins. DS3 has a
much higher granularity with 40500 total voxels, where the number of
layers is unchanged but the angular and radial binning is
50$\times$18. Both datasets have a threshold of 15.15~keV.  While this
is an unrealistic cut for practical applications, it provides a useful
challenge to high-dimensional generative networks covering a wide energy
range.

%%%%%%%%%%%%%%%%%%%%%%%%%%%%%%%%%%%%%%%%%%%%%%%%%%%
\subsection*{Preprocessing}
\label{sec:setup_pre}

We improve our training by including a series of preprocessing steps,
similar to previous studies~\cite{Krause:2022jna,Ernst:2023qvn,Diefenbacher:2023vsw,Amram:2023onf,Mikuni:2023tqg}. We split information on the
deposited energy from its distribution over voxels by introducing
energy ratios~\cite{Krause:2021ilc}
%-
\begin{align}
  u_0 = \frac{\sum_i E_i}{f \einc}
  \qqquad \text{and} \qqquad 
  u_i = \frac{E_i}{\sum_{j\ge i} E_j} \quad i=1,\ldots,44 \;,
\label{eq:enc_energy}
\end{align}
where $E_i$ refers to the total energy deposited in layer $i$, and
$f\in \mathbb{R}$ is a scale factor. The number of $u$-variables
matches the number of layers. With these variables extracted from a
given shower, we are free to normalize the voxel values by the energy
of their corresponding layer without losing any information. This
definition is analytically invertible, imposes energy conservation,
and ensures that the normalized voxels and each $u_{i>0}$ are always
in the range $[0,1]$. However, due to the calibration of the detector
response caused by the inactive material, $u_0$ can have values larger
than 1. We set $f=2.85$ in Eq.\eqref{eq:enc_energy}, to
rescale $u_0 \in[0,1]$.  All networks are conditioned on
$\einc$. This quantity is passed to the network after a log
transformation and a rescaling into the unit interval.

To train the autoencoders used for dimensionality reduction we do not
use any additional preprocessing steps. For the setup using the full input
space, we apply a logit transformation regularized by the parameter
$\alpha$ which rescales each input voxel $x$,
\begin{align}
x_\alpha &= (1-2\alpha)x + \alpha \in [\alpha,1-\alpha]
\qqquad \text{with} \quad \alpha=10^{-6}\notag \\
x' &= \log \frac{x_\alpha}{1-x_\alpha}\,.
\label{eq:alpha}
\end{align}
Finally, we calculate the mean and
the standard deviation of the training dataset and standardize each
feature. The postprocessing includes an additional step that rescales the sum of the generated voxels to ensure the correct normalization in each layer.

%%%%%%%%%%%%%%%%%%%%%%%%%%%%%%%%%%%%%%%%%%%%%%%%%%%
\section{CaloDREAM}
\label{sec:dream}

In CaloDREAM\footnote{The code used for this paper is publicly available at \href{https://github.com/heidelberg-hepml/calo_dreamer}{https://github.com/heidelberg-hepml/calo\_dreamer}}, we employ two generative networks, one energy network and one
shape network~\cite{Krause:2021ilc}. The energy network learns the energy-ratio features
conditioned on the incident energy, $p(u_i|\,\einc)$. The shape
network learns the conditional distribution for the voxels,
$p(x|\,\einc,u)$. The two networks are trained independently,
but are linked in the generative process. Specifically, to sample
showers given an incident energy, we follow the chain
\begin{align}
    u_{i} &\sim p_\phi(u_i|\einc) \notag \\ 
    x &\sim p_\theta(x|\einc, u) \; .
\end{align}
In this notation $\phi$ stands for the weights in the energy network 
and $\theta$ for the weights in the shape network.
Although the number of calorimeter layers is consistent across DS2 and
DS3 and the underlying showers are the same, we train separate energy networks for each dataset. The incident energy is always sampled from the known distribution in the datasets, as in Eq.\eqref{eq:einc-range}.

%%%%%%%%%%%%%%%%%%%%%%%%%%%%%%%%%%%%%%%%%%%%%%%%%%%
\subsection{Energy network --- Transfusion}
\label{sec:dream_energy}

Both of our generative networks use the Conditional Flow Matching
architecture~\cite{lipman2022flow}. It starts with the 
ordinary differential equation (ODE)
\begin{align}
  \frac{dx(t)}{dt} = v(x(t),t)
  \qquad \text{with} \qquad
  x\in\mathbb{R}^d \; ,
\label{eq:sample_ODE}
\end{align}
and a velocity field $v(x(t),t) \in \mathbb{R}^d$.
This time evolution can be related to the underlying density through the
continuity equation 
\begin{align}
\frac{\partial p(x,t)}{\partial t} + \nabla_x \left[ p(x,t) v(x,t) \right] = 0 \; .
\label{eq:continuity}
\end{align} 
The velocity field transforms the density $p(x,t)$ such
that
\begin{align}
 p(x,t) \to 
 \begin{cases}
  \normal(x;0,1) \quad & t \to 0 \\
  \pd(x)  \quad & t \to 1  \;.
\end{cases} 
\label{eq:cfm_limits}
\end{align}
We can estimate the velocity field with $v_\phi(x(t),t)$. In this case we can sample the data distribution from 
Gaussian random numbers, tracing the trajectory using any ODE
solver. Defining the training trajectories to be linear, the velocity network is optimized using a
simple MSE loss
\begin{align}
    \loss_\text{CFM} &= \XXLangle \left[ v_\phi((1-t)\epsilon+t x,t) - (x - \epsilon)\right]^2 \XXRangle_{t\sim U(0,1),\,\epsilon\sim \normal,\,x \sim \pd} \; .
\label{eq:didi_loss}
\end{align}
Conditional probability distributions can be learned by allowing
$v_\phi$ to depend on additional inputs.

%--------------------------------------------------------------
\begin{figure}[t]
    \centering
    \begin{tikzpicture}
[node distance=2cm, scale=0.65, every node/.style={transform shape}]

\node (Einc) [txt,font=\Large] {$E_{\text{inc}}$};

\node (emb_Einc) [expr, below of=Einc, yshift=-0.5cm, rotate=90, font=\Large]{Emb};

\node (TE) [transformer, below of=emb_Einc, yshift=-1.1cm, text width=4.5cm,
text depth=1.5cm, align=center, font=\Large] {Transformer-Encoder};
\node (TE_att) [attention, below of=TE, yshift=1.6cm, font=\Large] {Self-Attention};

\node (start) [txt, right of=Einc, xshift=3cm, font=\Large] {$0$};
\node (u0) [txt, right of=start, xshift=1cm, font=\Large] {$u_0$};
\node (u1) [txt, right of=u0,xshift=-0.2cm, font=\Large] {$...$};
\node (u2) [txt, right of=u1,xshift=-0.2cm, font=\Large] {$u_{43}$};

\node (emb_start) [expr, below of=start, yshift=-0.5cm, rotate=90, font=\Large]{Emb};
\node (emb_0) [expr, below of=u0, yshift=-0.5cm, rotate=90, font=\Large]{Emb};
\node (emb_1) [txt, below of=u1, yshift=-0.5cm, font=\Large] {$...$};
\node (emb_2) [expr, below of=u2, yshift=-0.5cm, rotate=90, font=\Large]{Emb};

\node (TD) [transformer_huge, right of=TE, xshift=6.3cm, yshift=-0.8cm, text width=4.7cm,
text depth=3.1cm, align=center, minimum height=4cm, font=\Large] {Transformer-Decoder};
\node (TD_att) [attention_huge, right of=TE_att, xshift=6.3cm, font=\Large] {Masked Self-Attention};
\node (TD_crossatt) [attention_huge, below of=TD_att, yshift=0.4cm, font=\Large] {Cross-Attention};

% \node (t) [txt, right of=u2_t, xshift=2cm] {$t$};
\node (inn_start_b) [small_cinn_black, below of=start, yshift=-10.5cm, rotate=90, font=\Large]{cINN};
\node (inn_start) [small_cinn, below of=start, yshift=-10.5cm, rotate=90, font=\Large]{CFM};
\node (inn0_b) [small_cinn_black, below of=u0, yshift=-10.5cm, rotate=90, font=\Large]{cINN};
\node (inn0) [small_cinn, below of=u0, yshift=-10.5cm, rotate=90, font=\Large]{CFM};
\node (inn1) [txt, below of=u1, yshift=-10.5cm, font=\Large]{$...$};

\node (inn2_b) [small_cinn_black, below of=u2, yshift=-10.5cm, rotate=90, font=\Large]{cINN};
\node (inn2) [small_cinn, below of=u2, yshift=-10.5cm, rotate=90, font=\Large]{CFM};

\node (u0_t) [txt, left of=inn_start, xshift=0.8cm,yshift=2cm, font=\Large] {$u_0(t)$,$t$};
\node (u1_t) [txt, left of=inn0, xshift=0.8cm, yshift=2cm,font=\Large] {$u_1(t)$,$t$};
 % \node (u2_t) [txt, below of=u1, yshift=-8.5cm] {...};
\node (u3_t) [txt, left of=inn2, xshift=0.8cm,yshift=2cm, font=\Large] {$u_{44}(t)$,$t$};

%\node (linear) [attention, below of=TD, yshift=-2.2cm, fill=Rcolor]{Linear Layer};

\node (prob0) [txt, below of=inn_start, yshift=-0.5cm,font=\Large]{$\Big( v_\phi(u_0(t), c_0, t),$ };
\node (prob1) [txt, below of=inn0, yshift=-0.545cm, xshift=0.2cm, font=\Large]{$ \; v_\phi (u_1(t),c_1,t),$};
\node (prob2) [txt, below of=inn1, yshift=-0.5cm,font=\Large, xshift=0.3cm]{$... \;$};
\node (prob6) [txt, below of=inn2, yshift=-0.5cm,xshift=0.7 cm, font=\Large]{$, \; v_\phi(u_{44}(t),c_{44}, t) \Big)$};
\node (prob) [txt, left of=prob0, xshift=-1.7cm,font=\Large]{$v_\text{full}(u(t), t, E_{\text{inc}}) = \;$};

\draw [arrow, color=black] (Einc.south) -- (emb_Einc.east);
\draw [arrow, color=black] (emb_Einc.west) -- (TE.north -| emb_Einc.west);

\draw [arrow, color=black] (TE.south -| emb_Einc.west) --  ([yshift=-1cm]TE.south -| emb_Einc.west) -- ([yshift=-1cm]TE.south -| TD.west) ;

\draw [arrow, color=black] (start.south) -- (emb_start.east);
\draw [arrow, color=black] (u0.south) -- (emb_0.east);
\draw [arrow, color=black] (u2.south) -- (emb_2.east);
\draw [arrow, color=black] (emb_start.west) -- (TD.north -| emb_start.west);
\draw [arrow, color=black] (emb_0.west) -- (TD.north -| emb_0.west);
\draw [arrow, color=black] (emb_2.west) -- (TD.north -| emb_2.west);
% (A) (B);

\draw [arrow, color=black] (TD.south -| emb_start.west)  -- node [text width=1.5cm, pos=0.3, font=\Large, right] {$c_{0}$} (inn_start.east -| emb_start.west);
\draw [arrow, color=black] (TD.south -| emb_0.west)  -- node [text width=1.5cm, pos=0.3, font=\Large, right] {$c_{1}$} (inn0.east -| emb_0.west);
\draw [arrow, color=black] (TD.south -| emb_2.west)  -- node [text width=1.5cm,pos=0.3, font=\Large, right] {$c_{44}$}  (inn2.east -| emb_2.west);

\draw [arrow, color=black] (inn_start.west -| emb_start.west) -- (prob0.north -| emb_start.west);
\draw [arrow, color=black] (inn0.west -| emb_0.west) -- (prob2.north -| emb_0.west);
\draw [arrow, color=black] (inn2.west -| emb_2.west) -- (prob6.north -| emb_2.west);

\draw[arrow, color=black](u0_t.south)--(inn_start.north -| u0_t.south) -- (inn_start.north);
\draw[arrow, color=black](u1_t.south)--(inn0.north -|u1_t.south) -- (inn0.north);
\draw[arrow, color=black](u3_t.south)--(inn2.north -| u3_t.south) -- (inn2.north);

\end{tikzpicture}
    \caption{Schematic diagram of the autoregressive Transfusion network~\cite{Heimel:2023mvw} used in our energy network.}
    \label{fig:artransfusion-diagram}
\end{figure}
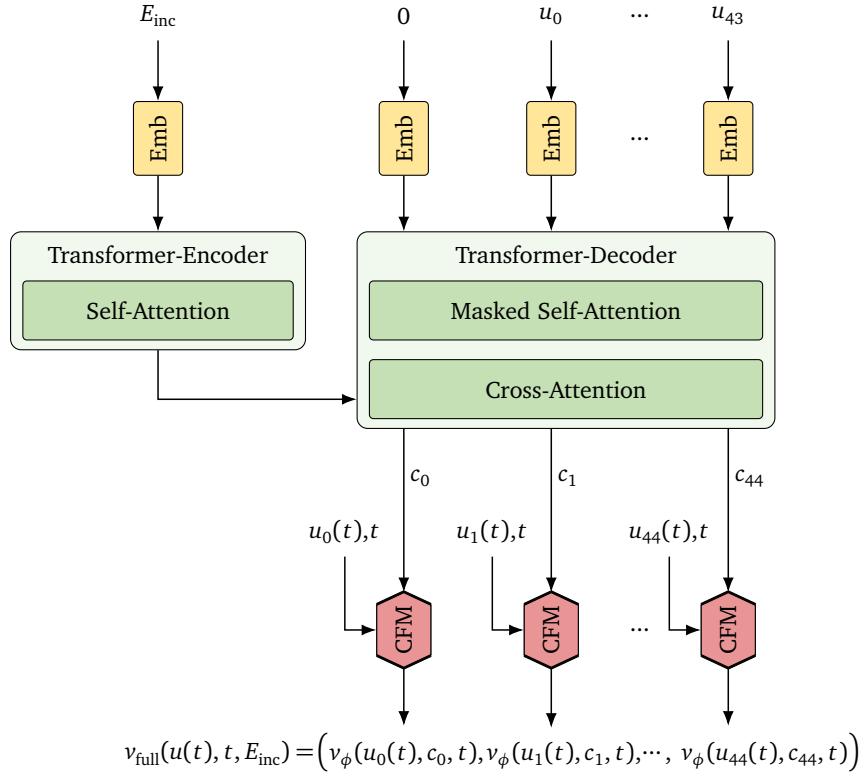
%--------------------------------------------------------------

For the energy network, we exploit the causal nature of the energy deposition 
in layers using an autoregressive transfusion architecture~\cite{Heimel:2023mvw},
as visualized in Fig~\ref{fig:artransfusion-diagram}.
We start by embedding $\einc$ as our one-dimensional condition
and the $u$-vector. For the $u$, this is done by 
concatenating a one-hot encoded position vector and zero-padding. These embeddings are passed to the encoder and decoder 
of a transformer, respectively.
For the one-dimensional condition the encoder's self-attention reduces to 
a trivial $1 \times 1$ matrix. For the decoder we mask our self-attention with an 
upper triangle matrix, to keep the autoregressive conditioning. Afterward,
we apply a cross-attention between the encoder and decoder outputs. 
The transformer 
outputs the vectors $c_0, \dots,\; c_{44}$, encoding the 
incident energy and previous energy ratios,
\begin{align}
    c_i &= \begin{cases}
       c_i(u_0, \dots, u_{i-1}, E_{\text{inc}}) \qqquad & i>0 \\
      c_i(E_{\text{inc}}) \qqquad & i=0  \; .
    \end{cases}
\end{align}
For generation, we use a single dense CFM network $v_\phi$, with the inputs 
time $t$, embedding 
$c_i$, and the point on the diffusion trajectory $u_i(t)$. This network is
evaluated 45 times to predict each component of the
velocity field individually,
\begin{align}
    v_{\text{full}}(u(t), t, \einc) &= \left( v_\phi(u_0(t),c_0,t), \dots, v_\phi(u_{44}(t),c_{44}, t) \right)
\label{eq:u_cfm}
\end{align}
During training, we can evaluate the contribution of each $u_i$ to the
loss in parallel, whereas sampling requires us to iteratively predict
the $u_i$ layer by layer. The hyperparameters of the transfusion network 
are given in Tab.~\ref{tab:energy_model_params}.

%%%%%%%%%%%%%%%%%%%%%%%%%%%%%%%%%%%%%%%%%%%%%%%%%%%
\subsection{Shape network --- Vision Transformer}
\label{sec:dream_shape}

%--------------------------------------------------------------
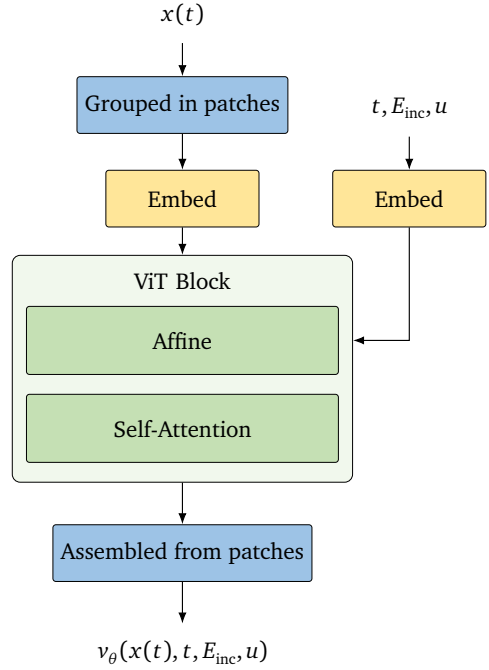
\begin{figure}[t]
    \centering
    \scalebox{0.75}{\begin{tikzpicture}

% Transformer body
\node (block) [transformer, text width=4cm,
text depth=3.1cm, align=center, minimum height=4cm,font=\large] {ViT Block};
\node (block_aff) [attention, above of=block, yshift=-0.50cm, font=\large] {Affine};
\node (block_atn) [attention, below of=block, yshift=-0.05cm, font=\large] {Self-Attention};

% Embedding layers
\node (embed_x) [expr, above of=block, yshift=2cm, xshift=0cm, minimum width=2.7cm, font=\large]{Embed};
\node (embed_c) [expr, above of=block, yshift=2cm, xshift=4cm, minimum width=2.7cm, font=\large]{Embed};
\draw [arrow, color=black] (embed_x.south) -- (block.north);
\draw [arrow, color=black] (embed_c.south) |- ([yshift=0.5cm]block.east);

% Encoding
\node (patch_x) [expr, above of=embed_x, yshift=0.65cm, inner sep=4pt, fill=Bcolor, minimum width=2.7cm, font=\large]{Grouped in patches};
\draw [arrow, color=black] (patch_x.south) -- (embed_x.north);

% Inputs
\node (x) [txt, above of=patch_x, yshift=0.6cm, font=\large]{$x(t)$};
\node (c) [txt, above of=embed_c, yshift=0.6cm, font=\large]{$t, E_\text{inc}, u$};
\draw [arrow, color=black] (x.south) -- (patch_x.north);
\draw [arrow, color=black] (c.south) -- (embed_c.north);

% Output
\node (vector) [txt, below of=block, yshift=-4cm, font=\large]{$v_\theta(x(t), t, E_\text{inc}, u)$};
\node (unpatch_v) [expr, below of=block, inner sep=4pt, yshift=-2.25cm, fill=Bcolor, minimum width=2.7cm, font=\large]{Assembled from patches};
\draw [arrow, color=black] (block.south) -- (unpatch_v.north);
\draw [arrow, color=black] (unpatch_v.south) -- (vector.north);

\end{tikzpicture}}
    \caption{Schematic diagram of the vision transformer (ViT)~\cite{Peebles2022ScalableDM} used in our shape network.}
    \label{fig:vit-diagram}
\end{figure}
%--------------------------------------------------------------

For the shape network, we use a 3-dimensional vision transformer (ViT) to learn the conditional velocity field $v_\theta(x(t), t, E_\text{inc}, u)$. The architecture is
inspired by Ref~\cite{Peebles2022ScalableDM} and illustrated in 
Fig.~\ref{fig:vit-diagram}. It divides the calorimeter 
into non-overlapping groups of voxels, so-called patches, which are
embedded using a shared linear layer and passed to a sequence of transformer blocks. Each block consists of a multi-headed self-attention and a dense network that transforms the 
patch features. To break the permutation symmetry among patches, we add a learnable position encoding to the patch embeddings prior to the first attention block. After the last block, a linear layer projects the processed patch features into the original patch dimensions, where each entry represents a diffusion velocity. Finally, the patches are reassembled into the calorimeter shape.

The network uses a joint embedding for the conditional inputs, $t,E_\text{inc}$ and $u$. The time and energy coordinates are embedded with separate dense networks, then summed into a single condition vector. The attention blocks incorporate this condition via affine transformations with shift and scale $a,b\in\R$ and an additional rescaling factor $\gamma\in\R$ learned by dense layers. These are applied within each block, and also to the final projection layer. Concretely, the operation inside the ViT block is summarized by
\begin{align}
 x_{\text{h}} &= x + \gamma_{\text{h}} g_{\text{h}} (a_{\text{h}}x + b_{\text{h}}), \notag \\
 x_{\text{l}} &= x_{\text{h}} + \gamma_{\text{l}} g_{\text{l}} (a_{\text{l}}x_{\text{h}} + b_{\text{l}}),
\end{align}
where $g_{\text{h}}$ is the multi-head self-attention step and $g_{\text{l}}$ is the fully connected transformation. The hyperparameters of our transformer are given in Tab.~\ref{tab:vit_params}.

The scalability of this architecture is closely tied to the choice of patching. On the one hand, small patches result in 
high-dimensional attention matrices. While this gives a more expressive network, the large number of operations can become a limitation for highly-granular calorimeters. Conversely, a large patch size compresses many voxels into one object, implying a faster forward pass but at the expense of sample quality. In this case, an expanded embedding dimension is needed to keep the network flexibility fixed. We decide on specific patch sizes for DS2 and DS3 through manual exploration.

Usually, we train Bayesian versions~\cite{Bellagente:2021yyh} of all our generative networks, including
calorimeter showers~\cite{Ernst:2023qvn}. In this study, the networks learning DS2 and DS3 are so heavy in terms of operations, that
an increase by a factor two, to learn an uncertainty map over phase space, surpasses our typical 
training cost of 40 hours on a cutting-edge NVIDIA H100 GPU. In principle, Bayesian versions of all
 networks used in this study can be built and used to quantify limitations, for instance related to 
a lack of training data.

%%%%%%%%%%%%%%%%%%%%%%%%%%%%%%%%%%%%%%%%%%%%%%%%%%%
\subsection{Latent diffusion}
\label{sec:dream_lat}

%--------------------------------------------------------------
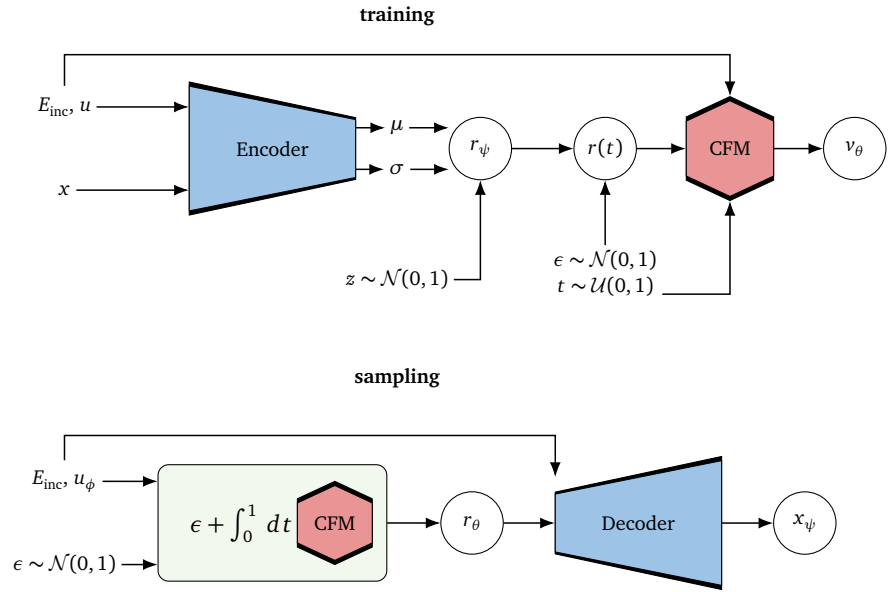
\begin{figure}[b!]
    \centering
    \begin{tikzpicture}[node distance=2cm, scale=0.55, every node/.style={transform shape}]
\node (text) [txt, font=\Large] {\textbf{training}};
\node (training) [bkg, below of = text, yshift=-2.2cm]{};
\node (conditions) [txt, below of= training, xshift=-8cm, yshift=4cm, font=\Large] {$E_{\text{inc}}$, $u$};
\node (input) [txt, below of = conditions, font=\Large]{$x$};
\node (encoder_b) [encoder_black,above of = conditions, yshift = 3cm, xshift= 1cm]{};
\node (encoder) [encoder,above of = conditions, yshift = 3cm, xshift= 1cm]{};
\node (encoder_txt) [txt, right of = conditions, yshift = -1cm, xshift= 3cm, font=\Large]{Encoder};
\node (mu) [txt, right of = encoder_txt, xshift=1cm, yshift=0.5cm, font=\Large]{$\mu$};
\node (sigma) [txt, right of = encoder_txt, xshift=1cm, yshift=-0.5cm, font=\Large]{$\sigma$};
\node (latent_input) [crc, right of = sigma, yshift=0.5cm, font=\Large]{$r_\psi$};
\node (z) [txt, below of = sigma, yshift=-0.65cm, font=\Large]{$z \sim \normal(0,1)$};
\node (r_t) [crc,right of= latent_input, xshift=1cm, font=\Large]{$r(t)$};
\node (cfm_b) [cinn_black,right of = r_t, xshift= 1cm]{};
\node (cfm) [cinn, right of = r_t, xshift= 1cm, font=\Large]{CFM};
\node (conditions_cfm) [txt,below of = r_t, yshift=-1cm, font=\Large]{ $\epsilon \sim \normal(0,1)$\\$t \sim \uniform (0,1)$}; 
\node (velocity) [crc, right of = cfm, xshift= 1cm, font=\Large]{$v_{\theta}$};

\node (text_samp)[txt, below of =training, yshift=-2.5cm, font=\Large]{\textbf{sampling}};
\node (sampling) [bkg, below of= training, yshift=-6cm, minimum height=6cm]{};
\node (conditions_samp) [txt, below of= sampling, xshift=-8cm, yshift=3cm, font=\Large] {$E_{\text{inc}}$, $u_\phi$};
\node (input_samp) [txt, below of = conditions_samp, font=\Large]{$\epsilon \sim \normal(0,1)$};
\node (integral) [transformer, minimum width=5.5cm, minimum height=2.8cm, font=\LARGE, right of = conditions_samp, yshift = -1cm, xshift= 3cm]{$\epsilon + \int_0^1 \;dt\;\;\;\;\;\;\;\;\;\;\;$};
\node (cfm_b_samp) [cinn_black,right of = conditions_samp, xshift=4.5cm ,yshift = -1cm,minimum height=2cm, minimum width=1.8cm]{};
\node (cfm_samp) [cinn, right of = conditions_samp, xshift=4.5cm ,yshift = -1cm, minimum height=1.8cm, minimum width=1.8cm, font=\Large]{CFM};
\node (r_model) [crc, right of =integral, xshift=2.8cm, font=\Large]{$r_{\theta}$};
\node (decoder_b) [encoder_black, rotate=180, below of= r_model, yshift=-2cm]{};
\node (decoder) [encoder, rotate=180, below of= r_model, yshift=-2cm]{};
\node (decoder_txt) [txt, right of= r_model, xshift=2cm, font=\Large]{Decoder};
\node (x_model) [crc, right of=decoder_txt, xshift=2cm, font=\Large]{$x_{\psi}$};

\draw [arrow, color=black](conditions.east) -- ([yshift=1cm]encoder_b.south);
\draw [arrow, color=black](conditions.north) -- ([yshift=1cm]cfm_b.north -| conditions.north) --([yshift=1cm] cfm_b.north) -- (cfm_b.north);
\draw [arrow, color=black](input.east) -- ([yshift=-1cm]encoder_b.south);
\draw [arrow, color=black]([yshift=0.5cm]encoder_b.north) -- (mu.west);
\draw [arrow, color=black]([yshift=-0.5cm]encoder_b.north) -- (sigma.west);
\draw [arrow, color=black] (mu.east) -- ([yshift=0.5cm]latent_input.west);
\draw [arrow, color=black] (sigma.east) -- ([yshift=-0.5cm]latent_input.west);
\draw [arrow, color=black] (z.east) -- (z.east -| latent_input.south)--(latent_input.south);
\draw [arrow, color=black](latent_input.east) -- (r_t.west);
\draw [arrow, color=black](conditions_cfm.north) -- (r_t.south);
\draw [arrow, color=black]([yshift=-0.5cm]conditions_cfm.east) -- ([yshift=-0.5cm]conditions_cfm.east -| cfm_b.south) -- (cfm_b.south);
\draw [arrow, color=black] (r_t.east) -- (cfm_b.west);
\draw [arrow, color=black] (cfm_b.east) -- (velocity.west);

\draw [arrow, color=black](conditions_samp.east) -- ([yshift=1cm]integral.west);
\draw [arrow, color=black](conditions_samp.north) -- ([yshift=1cm]decoder.east -| conditions_samp.north) --([yshift=1cm, xshift=-2cm] decoder.east) -- ([xshift=-2cm]decoder.east);
\draw[arrow, color=black](input_samp.east) -- ([yshift=-1cm]integral.west);
\draw[arrow, color=black](integral.east) -- (r_model.west);
\draw[arrow, color=black](r_model.east) -- (decoder.north);
\draw[arrow, color=black](decoder.south) -- (x_model.west);

\end{tikzpicture}
    \caption{Training (upper) and sampling (lower) with the latent diffusion network, using a variational autoencoder.}
\label{fig:latentdiffusion-diagram}
\end{figure}
%--------------------------------------------------------------

As the calorimeter granularity is increased from DS2 to DS3,
the computational requirements to train a network on the full voxel space also increase considerably
due to the larger number of patches.
This motivates a study of how 
the naive scaling may be avoided by a lower-dimensional latent representation. Starting from the detector geometry, a voxel-based representation of a shower defines a 
grid with fixed size and stores the deposited energy in each voxel.
This means a highly granular voxelization will produce a large fraction of zero voxels, but the
showers should define a
lower-dimensional manifold of the original phase space. 
Such a manifold can then be learned by an autoencoder~\cite{Cresswell:2022tof,Ernst:2023qvn,Liu:2024kvv}. 

We train a variational autoencoder with learnable parameters $\psi$.
The encoder outputs a latent parameter pair $(\mu, \sigma)$, which defines the latent variable $r = \mu + z\cdot\sigma$ with $z\sim \normal(0,1)$. The encoder distribution represents the phase 
space distributions over $x$ through
$p_\psi (r|x, u)$. For simplicity, in the following we drop the energy dependence in the encoder and decoder distributions. After sampling the latent variable, we minimize the learned likelihood of a Bernoulli decoder $p_\psi (x|r)$ represented by the reconstruction loss
\begin{align}
\loss_\text{VAE} = \Langle -\log p_\psi(x|r) \Rangle_{x\sim p_\text{data},r\sim p_\psi(r|x)} + \beta \Langle D_\text{KL} [p_\psi (r|x),\normal(0,1)]\Rangle_{x\sim p_\text{data}} \; .
\label{eq:bce_loss}
\end{align}
This choice of likelihood is possible since our preprocessing normalizes voxels 
into the range $[0,1]$. The reconstruction quality achieved in the autoencoder 
training places an upper bound on the
quality of a generative model trained in the corresponding latent space.

The KL-divergence term, with unit-Gaussian prior and a small weight $
\beta=10^{-6}$, is a regularization rather than a condition for a tractable 
latent space. It
encourages a smooth latent space, over which we train the 
generative network. Especially for DS3, an autoencoder trained without 
KL-regularization 
produces a sparse latent space with features mapped over several orders of magnitude.

The VAE consists of a series of convolutions, the last of 
which downsamples the data. This structure is mirrored in the decoder using 
ConvTranspose operations. 
As always, the energy conditions are encoded in a separate network 
and passed to the encoder and decoder. For a compressed latent space the ratio between the dimensionality of $x$ and $r$ defines the reduction factor $F$.
Rather than estimating the dimensionality of the datasets, we use a moderate, 
fixed reduction factor $F \simeq 2.5$ and 
a bottleneck with two channels. We do not expect the same reduction factor $F$ to be optimal for both datasets.
We provide more details on the autoencoder training in 
App.~\ref{app:ae}.

The trained autoencoder is used as a pre- and postprocessing step for the CFM as illustrated in 
Fig.~\ref{fig:latentdiffusion-diagram}.
Given the trained encoder distribution $p_\psi (r|x)$ the velocity field $v(r(t),t)$ imposes the boundary conditions
\begin{align}
 p(r,t) \to 
 \begin{cases}
  \normal(r; 0,1) \quad & t \to 0 \\
  p_\psi (r|x)  \quad & t \to 1,\, x\sim p_\text{data}  \,.
\end{cases}
\label{eq:lat_cfm_limits}
\end{align}
The expensive sampling then uses the lower-dimensional latent space and  
yields samples $r$ from the learned manifold. Finally, the phase space configurations are provided by the deterministic decoder $D_\psi(r)$. Here we summarize the sampling procedure, including the energy dependence, as three sequential steps:
\begin{align}
\label{eq:decoder}
u &\sim p_\phi(u|\einc) \notag \\
r &\sim p_\theta(r,1|u, \einc) \\
x &= D_\psi(r, u, \einc) \notag
\end{align}
All network hyperparameters and the main training parameters are
given in App.~\ref{app:hyper}.

%%%%%%%%%%%%%%%%%%%%%%%%%%%%%%%%%%%%%%%%%%%%%%%%%%%
\subsection{Bespoke samplers}
\label{sec:dream_bespoke}

A potential drawback of CFM networks is their slower sampling than, for instance, 
normalizing 
flows with coupling layers~\cite{Ernst:2023qvn}. This stems from the numerical integration of the ODE in 
Eq.\eqref{eq:sample_ODE}. Depending on the complexity of the target distribution, a standard ODE solver requires $\mathcal{O}(100)$ 
steps to achieve high-fidelity samples, each consisting of at least one forward pass of the neural network.

One method to overcome this slow inference is distillation~\cite{salimans2022progressive, pmlr-v202-song23a, Buhmann:2023kdg, Mikuni:2023tqg}, which 
aims to predict the sampling trajectory at only a handful of intermediate points, or even at 
the terminus in a single step. This requires fine-tuning the network weights using 
additional training time, in some cases even additional training data. Further, since 
the weights of the network itself are updated, consistency is not strictly guaranteed and 
we can end up sampling from a different distribution than was originally learned.

An alternative approach is to keep the network fixed and consider alternative structures for 
the ODE solver. Reference~\cite{Jiang:2024ohg} provides a comparison of various training-free solvers in 
the context of calorimeter simulations. 
While training-free approaches are the least costly, they are not task-specific and 
therefore unlikely to be optimal. 
However, there exists a trainable family of ODE solvers that can be optimized to a given vector 
field $v_\theta$ without excessive additional training~\cite{Shaul2023BespokeSF,Shaul2024BespokeNS}. 
Such Bespoke Non-Stationary (BNS) solvers parameterize the steps along the flow trajectory. Starting from an initial state $x_0$, and a time discretization $0=t_0<t_i<t_N=1$, the $i$\textsuperscript{th} integration step is
\begin{align}
    \label{eq:bns-step}
    x_{i+1} = a_i x_0 + b_i\cdot V_i
    \quad \text{with} \quad 
    a_i &\in\mathbb{R}, \; 
    b_i\in\mathbb{R}^{i+1} \notag \\
    V_i &= [v_\theta(x_0, t_0), \cdots, v_\theta(x_i, t_i)]\in \mathbb{R}^{(i+1)\times d}  \; ,
\end{align}
where we again suppress the energy dependence of $v_\theta$. By appropriately caching the velocities, each step requires just one evaluation of the network. In total, an $N$-step BNS solver has $N(N+5)/2+1$ learnable parameters: $a_i$, $b_i$ and the $t_i$ not fixed by the boundary 
conditions. Since this is typically orders of magnitude fewer than the network $v_\theta$, optimizing the solver requires a fraction of the computation time needed to train the vector field itself. Non-stationary solvers 
encompass a large family of ODE solvers, including the Runge-Kutta (RK) methods. Euler's method, i.e. first order RK, corresponds to taking $a_i=1$,$b_{ij}=1/N$ and evenly-spaced $t_i$.

Bespoke solvers can be trained by comparing the bespoke trajectory to a precisely-computed reference $x_\text{ref}(t)$, given an initial state $x_0$ sampled from the CFM latent distribution. Here we define two options. First, the global truncation error measures the deviation between the final states of the solvers
\begin{align}
    \label{eq:bns-global}
    \loss_\text{GTE} = \XLangle [x_\text{ref}(1) -  x_{N}]^2 \XRangle_{x_0\sim\mathcal{N}} \; ,
\end{align}
where $x_N$ is computed by iterating Eq.\eqref{eq:bns-step} starting from $x_0$. The 
local truncation error instead measures the discrepancy at each step,
\begin{align}
    \label{eq:bns-local}
    \loss_\text{LTE} 
    = \XLangle \sum_{i=0}^{N-1} \left[x_\text{ref}(t_{i+1}) - (a_i x_0 + b_i\cdot V_{\text{ref},i})\right]^2 \XRangle_{x_0\sim\mathcal{N}} \, ,
\end{align}
where $V_{\text{ref},i}$ is defined as in Eq.\eqref{eq:bns-step}, but with velocities evaluated on the reference trajectory.

Although we use CFM for both our shape and energy networks, we only study BNS solvers for the shape network. For training a BNS solver, we initialize it to the Euler method. At each iteration, we sample an $x_0$ batch from the unit Gaussian and a batch of conditions from the energy network. A precise solver is then used to generate the reference trajectory $x_\text{ref}(t)$ which enters the loss. Note that the shape model parameters $\theta$ are frozen during training. The complete list of hyperparameters are given in App.~\ref{app:hyper}.

%%%%%%%%%%%%%%%%%%%%%%%%%%%%%%%%%%%%%%%%%%%%%%%%%%%
\section{Results}
\label{sec:res}

%%%%%%%%%%%%%%%%%%%%%%%%%%%%%%%%%%%%%%%%%%%%%%%%%%%
\subsection{Layer energies}
\label{sec:res_energies}

%--------------------------------------------------------------
\begin{figure}
    \includegraphics[width=0.32\textwidth]{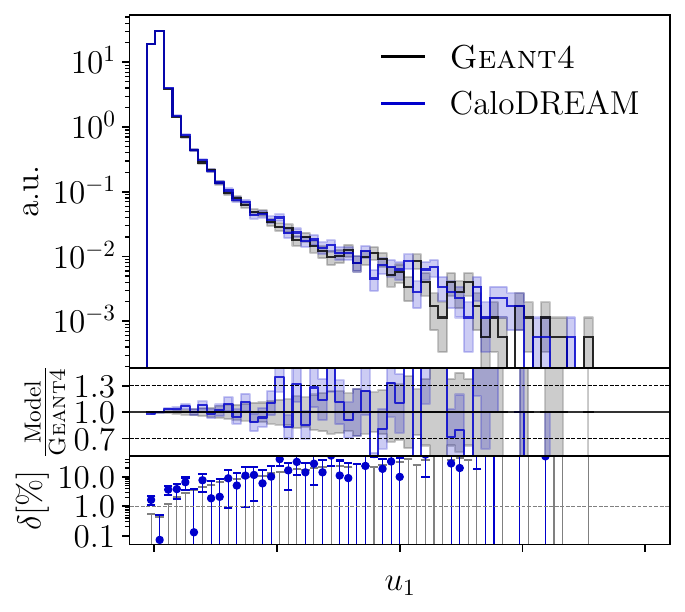}    
    \includegraphics[width=0.32\textwidth]{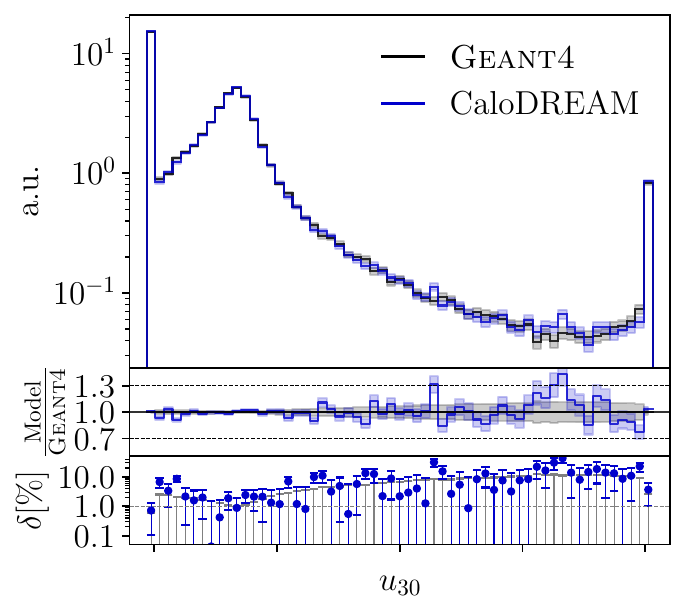}
    \includegraphics[width=0.32\textwidth]{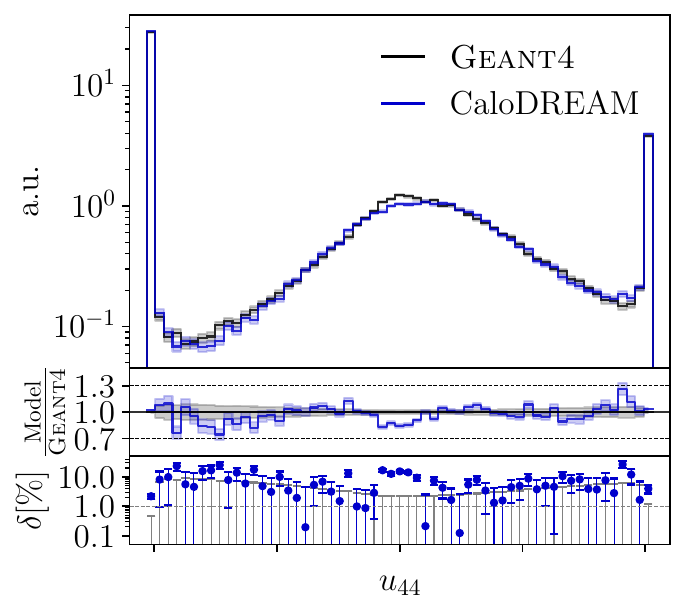}   
    \caption{Distributions of selected $u$-features in DS2 from 
      the CaloDREAM energy network (blue) compared to 
      truth (grey). The error bars in all feature distributions in this paper
      show the statistics of the respective datasets.}
    \label{fig:u_dists}
\end{figure}
%--------------------------------------------------------------

In Fig.~\ref{fig:u_dists} we compare samples generated from the energy network 
with the truth for a selection of normalized layer energies $u_i$. 
The transfusion network indeed generates high-quality distributions, with 
errors comparable to the statistical uncertainties in the 
test data. The distributions for $u_{i>40}$ are the most difficult to 
model, since the majority of showers lie in the sharp peaks at zero 
or one. These are zero-width peaks corresponding to showers that end at the given layer, leading to a one, or end before or skip the layer, leading to a zero.

We find that our autoregressive setup is particularly effective in 
faithfully mapping regions close to these peaks. As a quantitative performance 
measure, we train a classifier to distinguish the $u$'s defined by our 
energy network from the \geant truth, obtaining AUC scores around 0.51 on an independent test set.
The comparison in terms of layer energy is shown in Fig.~\ref{fig:ds2_full}.
The factorization procedure allows us to use the same energy network for the ViT and the laViT, effectively generating statistically-identical layer energy distributions.

%%%%%%%%%%%%%%%%%%%%%%%%%%%%%%%%%%%%%%%%%%%%%%%%%%%
\subsection{DS2 showers}
\label{sec:res_ds2}

Given the learned layer energies, we use the shape networks described in
Sec.~\ref{sec:dream_shape} to generate the actual calorimeter showers
over the voxels.
First, we evaluate the distribution of energy depositions per 
layer by looking at shape observables, like 
the center of energy of the shower and its width in the 
$\phi$ and $\eta$ directions,
\begin{align}
\langle \xi \rangle 
&= \frac{\xi\cdot x}{\sum_i x_i} \notag \\
\sigma_{\langle \xi \rangle} &= \sqrt{\frac{\xi^2\cdot x}{\sum_i x_i} - \langle \xi \rangle^2}
\qqquad \text{for} \qqquad \xi \in \{\eta, \phi\} \; .
\end{align}
Here $x_i$ is the energy deposition in a single voxel and the sum
runs over the voxels in a layer. 

\begin{figure}[t]
    \includegraphics[width=0.33\textwidth, page=21]{figs/d2_full_latent_vit/E_layer_dataset_2.pdf} 
    \includegraphics[width=0.33\textwidth, page=1]{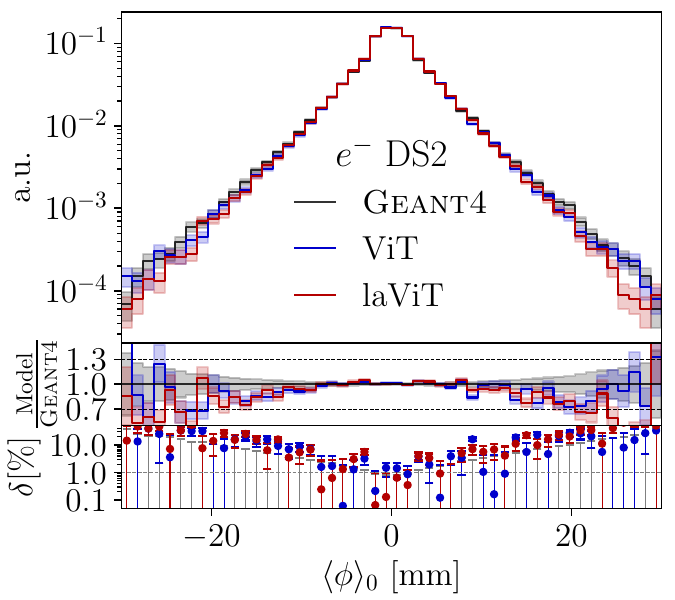} 
    \includegraphics[width=0.33\textwidth, page=11]{figs/d2_full_latent_vit/WidthPhi_layer_dataset_2.pdf} \\
    \includegraphics[width=0.33\textwidth, page=2]{figs/d2_full_latent_vit/Weighted_Depth_ring_dataset_2_groups_1.pdf}
    \includegraphics[width=0.33\textwidth]{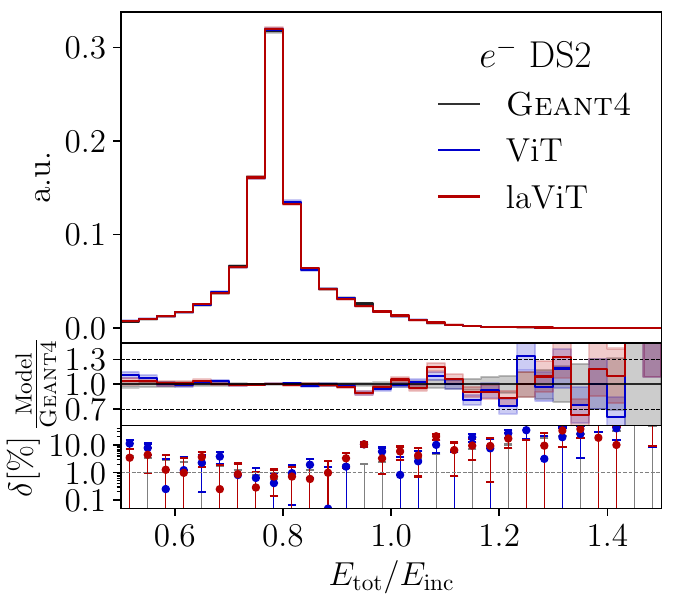} 
    \includegraphics[width=0.33\textwidth, page=1]{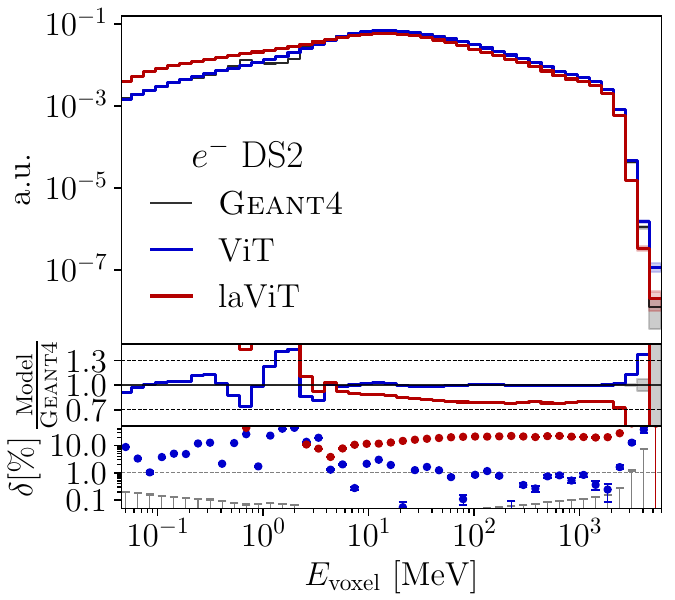} 
    \caption{Selection of high-level features for DS2. The first row shows 
    features for individual layers, the second row the combination of layers.}
    \label{fig:ds2_full}
\end{figure}

In the first row of Fig.~\ref{fig:ds2_full} we compare a set of 
layer-wise
distributions from the networks trained in the full space and in the
latent representation to the test data truth. We start with the 
energy deposited in layer 20, where for $E_{20} > 10$~MeV the 
full-dimensional vision transformer (ViT) as well as the latent-diffusion 
counterpart (laViT) 
agree with the truth at the level of a few per-cent, as expected. Towards smaller energies 
we see a missing feature in both networks. Also in the two other shown distributions the ViT 
and laViT agree with each other and deviate from \geant only in regions with 
statistically limited training data.

The second row of
Fig.~\ref{fig:ds2_full} shows example
distributions probing the combination of layers.
In addition to the layer-wise shower shapes,
we calculate the mean shower depth
weighted by the energy deposition in each of the $N$ layers 
for slices in the radial direction,
\begin{align}
d_{r_j} = \frac{\sum_i^{N} k_i E_{i,r_j}}{E_{\text{tot},j}} \qquad 
\qquad r_j \in \{0, \ldots, |r|\}  \; .
\label{eq:sh_depth}
\end{align}
Here $E_{i,r_j}$ is the average
energy deposition in slice $r_j$, and
$E_{\text{tot},j}$ is the total energy deposition in the selected slice. 
Slices in the angular direction are less interesting to calculate due to the rotational invariance of the showers.
This observable highlights a small deviation for both networks from the 
reference for showers with maximum depth of five layers not captured by
the layer-wise high-level features.

%-------------------------------------------------
%-------------------------------------------------

Also combining layer-wise information, we show the total energy deposited in the calorimeter $E_\text{tot}$ normalized by the incident energy and the full voxel distribution across the entire calorimeter $E_\text{voxel}$.
The total shower energy relative to the incident energy is reproduced very
well by both networks since this information is coming from the energy network. However for the voxel energies
only the full-dimensional network captures the low-energy regime, 
whereas the latent model overestimates this regime and in turn
shifts down the prediction for larger energies because of the 
normalization of the curve. This is the only noteworthy 
shortcoming of the laViT compared to the ViT that we find.

%-------------------------------------------------
\begin{figure}[b!]
    \includegraphics[width=0.33\textwidth]{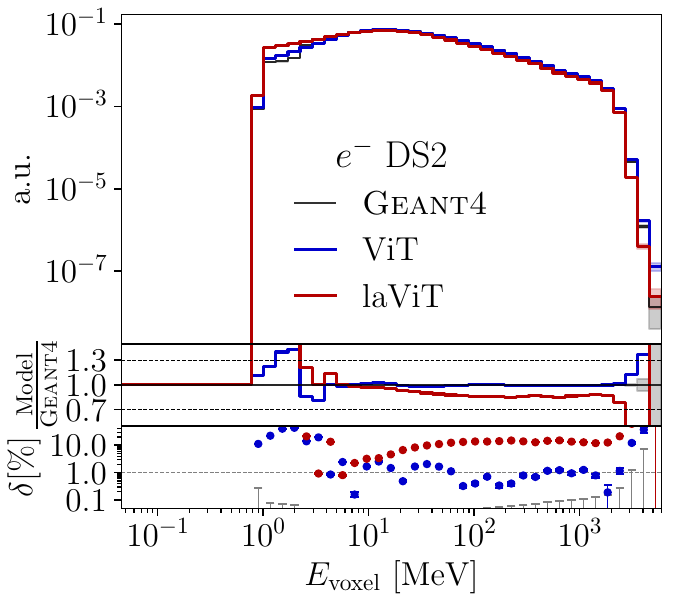} 
    \includegraphics[width=0.33\textwidth, page=11]{figs/d2_full_latent_vit/Sparsity_layer_dataset_2.pdf} 
    \includegraphics[width=0.33\textwidth, page=11]{figs/d2_full_latent_vit/Sparsity_layer_dataset_2_wcut.pdf} 
    \caption{Effect of an additional threshold $E>1$~MeV on DS2; we show the shower energy and the 
    sparsities without and with threshold cut.}
    \label{fig:ds2_sp}
\end{figure}
%-------------------------------------------------

Following up on the problem raised by the last panel in Fig.~\ref{fig:ds2_full}, 
we focus on the (latent) 
description with low-energy voxels. In Fig.~\ref{fig:ds2_sp} we 
again compare the two network predictions with the truth, but applying 
an additional threshold cut of
\begin{align}
 E_\text{voxel} > 1~\mev \; .
\end{align}
After this cut, the agreement of the laViT prediction with the 
full ViT and the truth improves significantly. We checked that this cut has only a limited impact on the total energy deposition $E_\text{tot}$. Slight deviations 
are limited to the threshold region $E_\text{voxel} \lesssim 5$~GeV. The reason 
can be seen in the sparsity distributions for instance of layer~10,
$\lambda_{10}$.
The laViT network generates a sizable number of showers with energy depositions
everywhere, leading to a peak at zero sparsity. This
failure mode is already present in the autoencoder reconstruction as described in App.~\ref{app:ae}.
Because of their low energy, these
contributions do not affect the other high-level observables or
the learned physics patterns of the showers.

%%%%%%%%%%%%%%%%%%%%%%%%%%%%%%%%%%%%%%%%%%%%%%%%%%%
\subsection{DS3 showers}
\label{sec:res_ds3}

%-------------------------------------------------
\begin{figure}
    \includegraphics[width=0.33\textwidth, page=21]{figs/d3_full_latent_vit/E_layer_dataset_3.pdf}
    \includegraphics[width=0.33\textwidth, page=1]{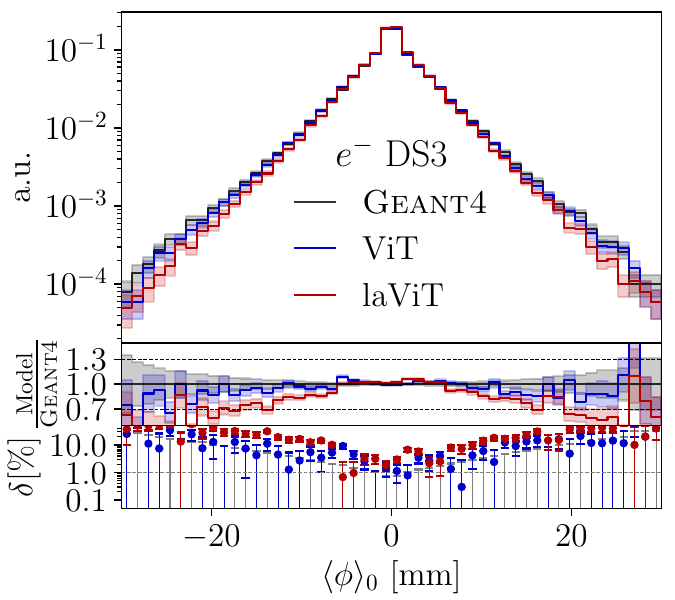}
    \includegraphics[width=0.33\textwidth, page=1]{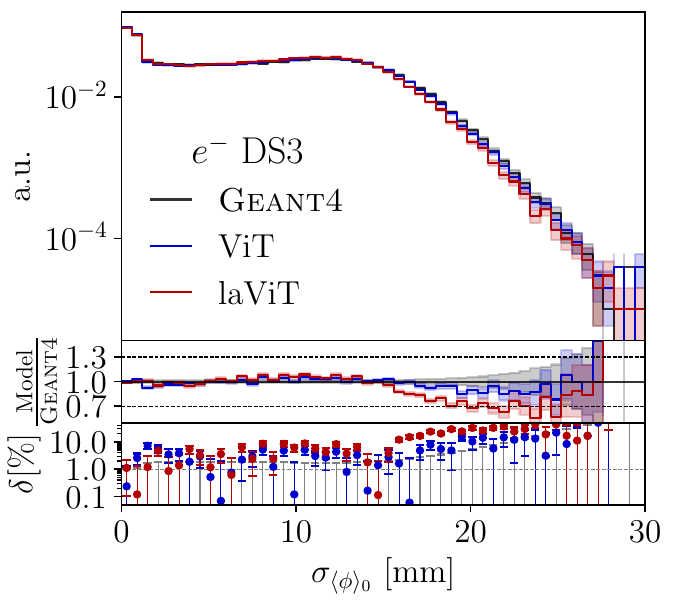}
 \\
     \includegraphics[width=0.33\textwidth, page=1]{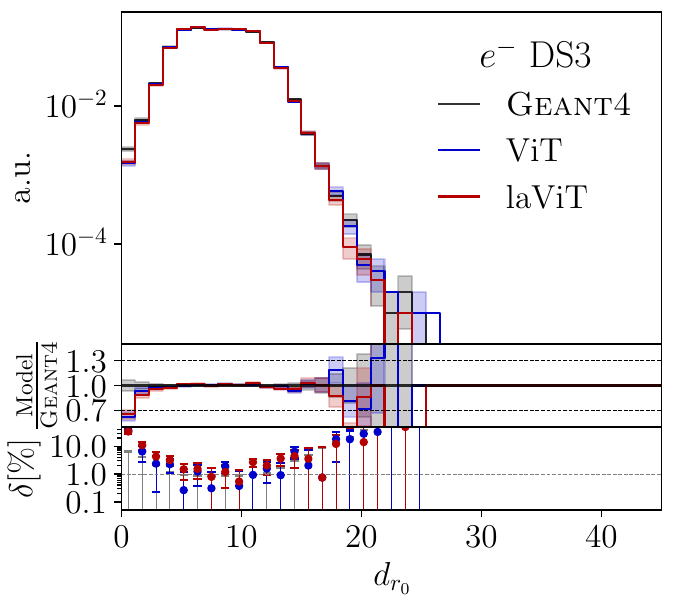}
    \includegraphics[width=0.33\textwidth]{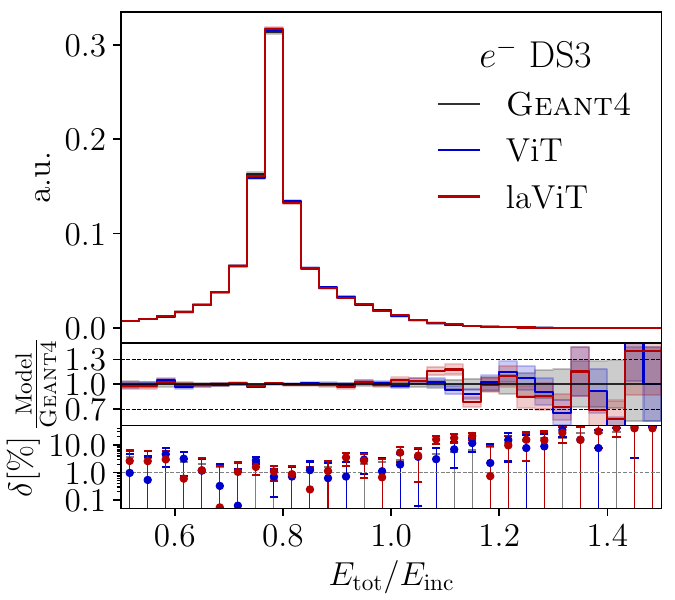}
    \includegraphics[width=0.33\textwidth]{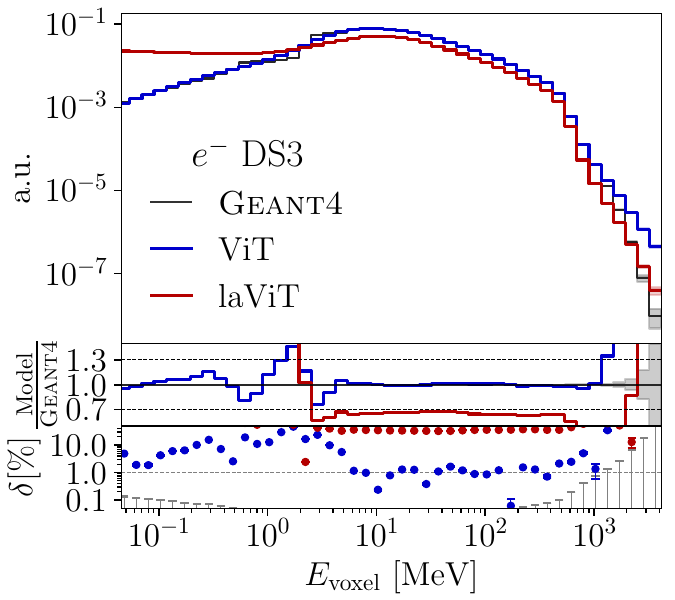} 
    \caption{Selection of high-level features for DS3. The first row shows
    features for individual layers, the second row the combination of layers. 
    All features correspond to the DS2 results shown in Fig.~\ref{fig:ds2_full}.}
    \label{fig:ds3_full}
\end{figure}
%-------------------------------------------------

The same analysis done for DS2 in Sec.~\ref{sec:res_ds3} we now 
repeat for DS3. This means we study the same shower energies and shower shapes, 
but from 40500 instead of 6480 voxels. A target
phase space of such large dimension is atypical for most LHC applications, and the key question is
whether the precision-generative architectures that have been successful on lower-dimensional phase 
spaces also give the necessary precision for high-dimensional phase spaces.
As a matter of fact, we know that this is not the case for standard 
normalizing flows or INNs~\cite{Ernst:2023qvn}, where the architectures 
have to be modified significantly to cope with higher resolution.

%-------------------------------------------------
\begin{figure}[t!]
    \includegraphics[width=0.33\textwidth]{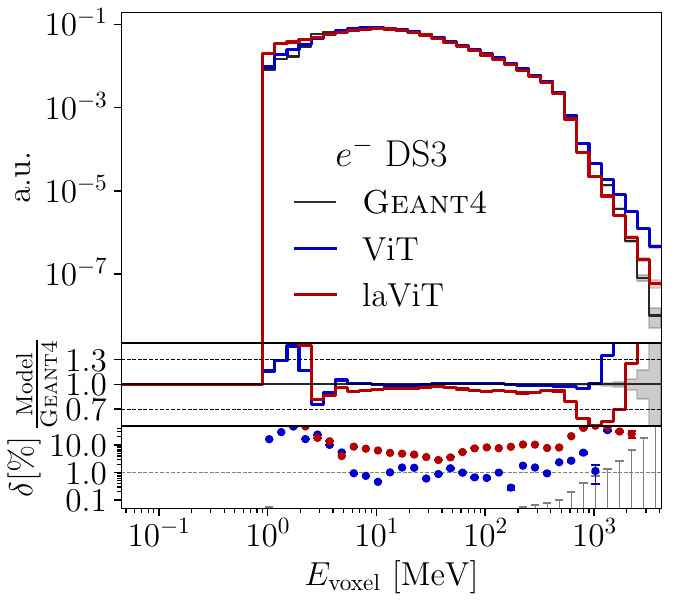} 
    \includegraphics[width=0.33\textwidth, page=11]{figs/d3_full_latent_vit/Sparsity_layer_dataset_3.pdf} 
    \includegraphics[width=0.33\textwidth, page=11]{figs/d3_full_latent_vit/Sparsity_layer_dataset_3_wcut.pdf}  
    \caption{Effect of an additional threshold $E>1$~MeV on DS3; we show the shower energy and the sparsities without and with threshold cut.
    All features correspond to the DS2 results shown in Fig.~\ref{fig:ds2_sp}.}
    \label{fig:ds3_sp}
\end{figure}
%-------------------------------------------------

In Fig.~\ref{fig:ds3_full} we again show a set of layer-wise features in 
the first row, observing extremely mild differences to the DS2 
results. Only the shower shapes from the laViT suffer slightly in 
regions with too little training data. For the multi-layer features 
in the second row, we also find the same results as for DS2, including the 
challenge in describing voxels with $E_\text{voxel} \lesssim 3$~GeV.

Understanding and targeting this challenge, we again show the voxel energy 
distribution and the sparsity after the threshold cut $E_\text{voxel} > 1$~MeV in 
Fig.~\ref{fig:ds3_sp}. For DS3 it turns out that after applying this cut the 
description of DS3 through the laViT network is excellent. The reason for this
is two-fold. Given the low energy bound we can reproduce with the latent model, 
a cut larger than this threshold completely adjusts the sparsity up to a specific 
value by removing the additional energy deposition of the latent model and the 
noisy components of \geant.
For both DS2 and DS3 the cut fixes the sparsity in $\lambda_{10}$ down to 
$\lambda_{10}\gtrsim 0.7$. However, for DS3 this is done by moving the peak at zero, 
while for DS2 the mass is moved from the intermediate sparsity.
This second difference comes from the dimensionalities of the two datasets, where
the fixed reduction factor has a stronger impact on DS2 due to the larger information 
loss in the bottleneck.

%%%%%%%%%%%%%%%%%%%%%%%%%%%%%%%%%%%%%%%%%%%%%%%%%%%
\subsection{Sampling efficiency}
\label{sec:sampling}

To demonstrate the performance of bespoke samplers, we compare the quality of showers produced by various solvers in terms of classifier tests. Classifiers trained to distinguish generated and true samples are an effective diagnostic tool since they capture failure modes in high-order correlations that are hidden in simple high-level distributions. As we will see in the following section, the phase space distribution of classifier scores can be used to search for and identify such failure modes. In this section, we only use the AUC as a simple, one-dimensional quality measure. 
The high-level classifier uses the layer-wise features but since we want sensitivity also to voxel-level correlations, we train a classifier on the low-level phase space as defined by the original voxels. 

\begin{figure}[t!]
    \includegraphics[width=\textwidth]{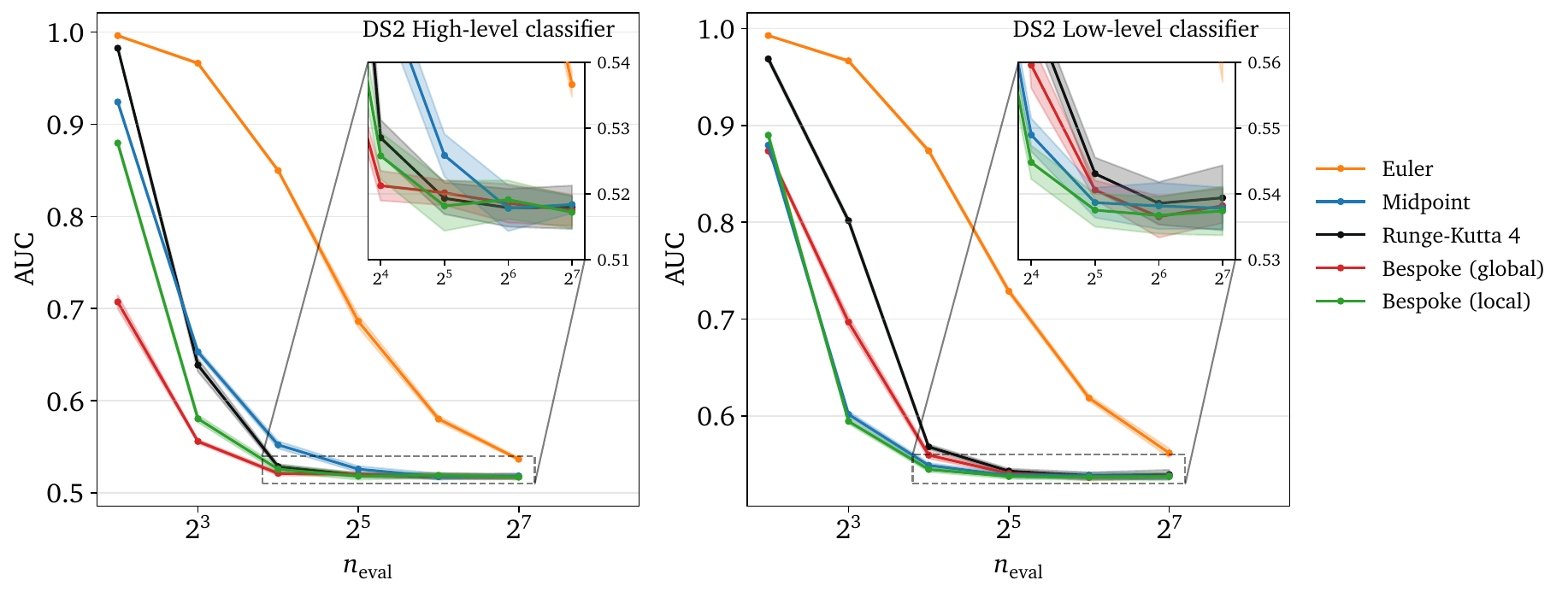}
    \caption{High-level (left) and low-level (right) classifier AUC scores on DS2 as a function of the number of function evaluations $n_\text{eval}$ for various ODE solvers. Samples of 100k reference and generated showers are used to train the classifier. Errors bands are taken as the standard deviation over 10 runs.}
    \label{fig:nfe_benchmark}
\end{figure}

For our comparison, we include three standard fixed-steps solvers: the Euler, Midpoint, and Runge-Kutta 4 methods. We also consider bespoke non-stationary solvers using either the global, Eq.\eqref{eq:bns-global}, or local, Eq.\eqref{eq:bns-local} truncation error as described in Sec.~\ref{sec:dream_bespoke}. Using each solver, we generate 100k showers from the DS2 ViT shape network. We train classifiers to distinguish these samples from the \geant reference set using the standard CaloChallenge pipeline.
In Fig.~\ref{fig:nfe_benchmark}, we plot the high-level (left) and low-level (right) AUC scores against the number of function evaluations $n_\text{eval}$ for each solver. Note that the Midpoint and RK4 methods respectively use 2 and 4 function evaluations per integration step. See App.~\ref{app:timing} for function evaluation timings of each network.

In both panels of the figure, we see that the Euler solver has a notably poor efficiency in terms of function evaluations. This indicates that the velocity field learned by the shape network has non-trivial curvature. Considering the remaining solvers, the sample quality essentially saturates by $n_\text{eval}=64$ and all non-Euler methods appear to have statistically-equal performance at this point. The bespoke samplers demonstrate the best retention in quality when looking toward smaller $n_\text{eval}$. In particular, the local BNS solver keeps an AUC below 0.6 for both classifiers even at 8 function evaluations. The global BNS solver achieves a large margin of improvement at $n_\text{eval}=4$ for the high-level classifier. The local bespoke solver also shows an advantage in the high-quality regime. Specifically, its AUC is already saturated for both solvers at 32 function evaluations. As such, in a resource-limited scenario the efficiency gains offered by bespoke solvers can be translated into improved sample quality.

It is interesting to note that the performance of a given solver can be significantly different between high- and low-level classifiers. This is evident in the reversed rankings of, for example, the two bespoke solvers in each panel. The global BNS solver favors performance on the high-level classifier, while the local BNS solver is best on the low-level classifier. A similar exchange can be seen among the Midpoint and RK4 solvers, with the former being close to optimal at low level.
%--------------------------------------------------------------
%--------------------------------------------------------------

%%%%%%%%%%%%%%%%%%%%%%%%%%%%%%%%%%%%%%%%%%%%%%%%%%%
\subsection{Performance}
\label{sec:res_class}

It is not trivial to test the overall performance of generative networks for calorimeter 
showers.
In the previous sections we evaluated the networks using simple one-dimensional histograms, as in Figs.~\ref{fig:ds2_full} and~\ref{fig:ds3_full}, or classifier AUC scores. 
%A simple way to probe physics captured in their showers is to 
%compare distributions of high-level features, as 
%shown in Figs.~\ref{fig:ds2_full} and~\ref{fig:ds3_full}. In the 
%ML4Jets CaloChallenge most of the high-level comparisons test the
%network performance by layer. Any
%mismodeling related to the shower depth will not be identified 
%from the standard histograms. 
A systematic approach to assess 
the quality of our generative networks, and a way to 
identify failure modes, is to examine the distribution of classifier predictions %train a classifier $D(x)$ 
over the phase space or feature space $x$~\cite{Das:2023ktd}.
%Following the 
%Neyman-Pearson lemma, 
A properly trained and calibrated classifier $C(x)$ learns the 
likelihood-ratio between the data and the generated distributions which, according to the Neyman-Pearson lemma, is the most powerful test statistic to discriminate between the two samples. This allows us to extract a correction
weight over phase space
\begin{align}
w(x) = \frac{C(x)}{1-C(x)} \approx \frac{p_\text{data}}{p_\text{model}}(x) \; ,
\label{eq:def_weights}
\end{align}
and to use the corresponding weight distributions as an
evaluation metric. The weights have to be evaluated on the training data and on 
the generated data, because failure modes appear as 
tails in one of the two distributions~\cite{Das:2023ktd}. To further 
analyze such failure modes, we can study 
showers with small or large
weights as a function of phase space, using the interpretable nature of 
phase spaces in particle physics. 

%--------------------------------------------------------------
\begin{figure}[t]
    \begin{minipage}[t!]{0.73\linewidth}\centering
    \includegraphics[width=0.495\textwidth]{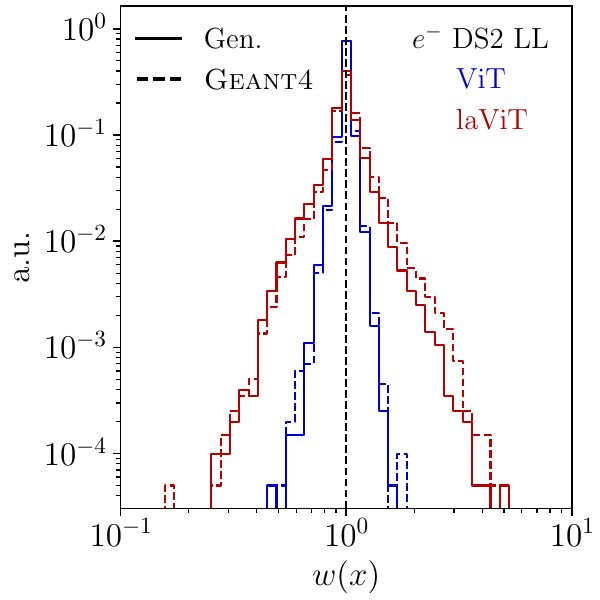}
    \includegraphics[width=0.495\textwidth]{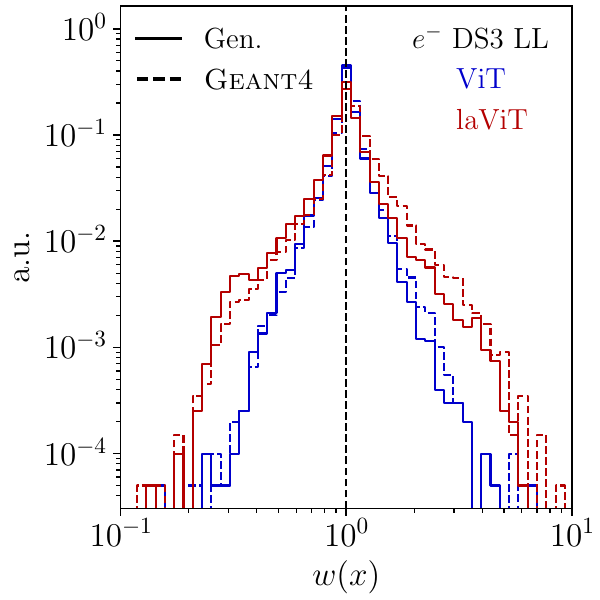}
    \end{minipage}
    \begin{minipage}[t!]{0.265\linewidth}
        \centering
        \begin{footnotesize}
        \begin{tabular}{lcccc}
            \toprule
             & \multicolumn{2}{c}{AUC (LL/HL)} \\ 
             \cmidrule{2-3}
             & DS2 \\
            \midrule
            ViT    & 0.535(3)/0.517(3) \\
            laViT  & 0.572(3)/0.536(2) \\ \midrule
            & DS3 & \\
            \midrule
            ViT    & 0.63(1)/0.525(3) \\
            laViT  & 0.62(1)/0.596(4) \\
            \bottomrule
        \end{tabular}
        \end{footnotesize}
    \end{minipage}
    \caption{Learned low-level (LL) classifier weight distributions for DS2 (left) and DS3 (right). We compare the full-dimensional ViT and the latent laViT results and, for each of them, show 
    weights for the generated sample and for a \geant test sample. The 
    table shows the AUC values for both high-level (HL) and low-level classifiers in each case.}
    \label{fig:weights}
\end{figure}
%--------------------------------------------------------------

In Fig.~\ref{fig:weights} we show the classifier 
weights from the low-level classifier for DS2 and for DS3.
We also include a table with the AUC scores of the high-level classifier trained on layer-wise features and the low-level classifier, where the ViT shows state-of-the-art results on DS2 and the high-level DS3.
The peaks of the weight distributions are nicely centered around $w=1$,
symmetric towards small and large (logarithmic) classifiers, and show no
significant difference between generated and training data. The weights for the
networks encoding the full phase space and the latent diffusion are different,
with a typical broadening of the distribution by a factor two around the 
peak and larger and less smooth tails. We still observe that the classifier misses the low-energetic noise affecting the sparsity and the 
voxel energy distributions.
Despite the simple nature of the neural network, a sequence of fully connected layers, the main result from this performance 
test is that the classifier identifies additional failure modes related to
the step from DS2 to DS3 and to the reduced latent space.
We expect these failure modes correspond to cross-layer features, since we observe a correlation between the classifier weights and the shower depth introduced in Sec.~\ref{sec:res_ds2}, and the high-level AUC is similar across the two datasets. Details of the neural network classifier are listed in App.~\ref{app:hyper}.

%%%%%%%%%%%%%%%%%%%%%%%%%%%%%%%%%%%%%%%%%%%%%%%%%%%
\section{Outlook}
\label{sec:outlook}

Calorimeter showers are one of the most exciting applications of 
modern generative networks in fundamental physics. Their specific 
challenge is the high dimensionality of the voxelized phase space, combined with 
extremely sparse data and an LHC-level precision requirement.
In our case, the CaloChallenge datasets~2 and~3 include up 
to 40k dimensions for the target phase space.

In this situation, diffusion networks allow us to go a step beyond 
standard normalizing flows. 
Our CaloDREAM architecture first factorizes the generation of detector
showers into an energy network and a shape network.
Both networks are trained using Conditional Flow Matching.
The former generates 
the layer energies using a transformer backbone with self-attention and 
cross-attention blocks.
For the latter, we use a 3-dimensional vision transformer, operating on 
patches of the target phase space.

For DS2 this combination of networks is a safe architecture choice, 
in the sense that it can be trained without problems and reproduces
all features, within layers and across layers, with high precision. 
We can use a VAE to reduce the dimensionality using latent diffusion.  
We find essentially no loss in performance, except for the reproduction of low-energy voxels
and, with it, sparsity, which can be improved by introducing 
an MeV-level energy threshold.
Because diffusion networks are 
slower than alternative generative networks, we use bespoke samplers to 
enhance their generation speed, at no cost of the precision
and improving the fidelity in case of limited resources.

For DS3 the performance of the CaloDREAM generators remains 
qualitatively the same, but the shape network reaches the limit 
in terms of available computation time. 
This is a motivation to again employ latent diffusion. We find excellent
performance of the latent diffusion architecture; with the right 
choice of energy thresholds even the 
sparsity distribution is reproduced correctly. 
However, irrespective of the dataset, further studies are needed to understand the effects of mapping the
distributions into real detectors with irregular geometries, more complex distributions from different incident particles, e.g. hadrons, and varying angle of impact. 

Our study shows that modern generative networks can be 
used to describe calorimeter showers in highly granular calorimeters.
When the number of phase space dimensions becomes 
very large and the data becomes sparse, a latent diffusion network
combined with an (autoregressive) transformer and bespoke sampling 
provides excellent benchmarks in speed and in precision.
We publish the generated samples together with the full set of high-level features in the Zenodo repository \href{https://doi.org/10.5281/zenodo.14413046}{10.5281/zenodo.14413046}.

\paragraph{Note added} A potentially similar approach, CaloDiT, also 
using a diffusion 
transformer to tackle calorimeter showers 
has been shown at ACAT 2024.

\section*{Acknowledgements}

We would like to thank Theo Heimel, Lorenz Vogel, and Anja Butter for 
extremely helpful discussions.
This research is supported 
through the KISS consortium (05D2022) funded by the German Federal Ministry of Education and Research BMBF 
in the ErUM-Data action plan,
by the Deutsche
Forschungsgemeinschaft (DFG, German Research Foundation) under grant
396021762 -- TRR~257: \textsl{Particle Physics Phenomenology after the
  Higgs Discovery}, and through Germany's Excellence Strategy
EXC~2181/1 -- 390900948 (the \textsl{Heidelberg STRUCTURES Excellence
  Cluster}). SPS
is supported by the BMBF Junior Group Generative Precision Networks for Particle Physics
(DLR 01IS22079). The authors acknowledge support by the state of Baden-W\"urttemberg
through bwHPC and the German Research Foundation (DFG) through grant no INST 39/963-1
FUGG (bwForCluster NEMO)

\clearpage
\appendix
%%%%%%%%%%%%%%%%%%%%%%%%%%%%%%%%%%%%%%%%%%%%%%%%%%%
\section{Further details}

\subsection{Hyperparameters}
\label{app:hyper}

%This section collects the fial parameters used for the networks
%presented in the previous
%sections. Tabs.~\ref{tab:energy_model_params},~\ref{tab:vit_params}
%shows the parameters for the energy network and the shape network of
%Sec.~\ref{sec:dream} respectively. Tab.~\ref{tab:cls_params}
%%contains the parameters used to train the classifier used for the 
%learned weight distributions of Fig.~\ref{fig:weights}.

%--------------------------------------------------------------
\begin{table}[h]
    \centering
    \begin{small} \begin{tabular}{lc} \toprule
        Parameter     & DS2 \& DS3  \\ \midrule
        % Param     & Choice (Dataset 2) & Choice (Dataset 3)\\ \midrule
        % \multirow{2}{*}{Param} & \multicolumn{2}{c}{Choice}\\
        % & Dataset 2 & Dataset 3 \\\midrule
        Epochs    & 500         \\
        LR sched. & cosine \\
        Max LR    & $10^{-3}$         \\
        Batch size & 4096 \\
        ODE solver  & Runge-Kutta 4 (50 steps) \\
        \midrule
        Network & transformer \\
        Dim embedding  &  64  \\
        Intermediate dim   & 1024 \\
        Num heads  & 4 \\
        Num layers   & 4 \\
        \midrule
        Network &  dense feed-forward\\
        Intermediate dim & 256 \\
        Num layers & 8\\
        Activation & SiLU\\
        \bottomrule
    \end{tabular} \end{small}
    \caption{Parameters for the autoregressive energy network in Sec.~\ref{sec:dream_energy}.}
    \label{tab:energy_model_params}
\end{table}
%--------------------------------------------------------------

%--------------------------------------------------------------
\begin{table}[h]
    \centering
    \begin{small} \begin{tabular}{lcccc} \toprule
                      & \multicolumn{2}{c}{ViT} & \multicolumn{2}{c}{laViT} \\ \midrule
        Parameter     & DS2 & DS3 & DS2 & DS3 \\ \midrule
        Patch size    &  (3, 16, 1) & (3, 5, 2) & (3, 1, 1) & (3, 2, 2) \\
        Embedding dimension   & 480 & 240 & 240 & 240 \\
        Attention heads  &  6 &  6 & 6 & 6 \\
        MLP hidden dimension & 1920 & 720 & 960 & 960 \\
        Blocks  &  6 &  6 & 10 & 10 \\
        % Dropout & 0.0 & 0.0 & 0.0 &\\
        \midrule
        epochs    & 800      &  600    & 800 &  400  \\
        batch size& 64       &  64     & 128 & 128 \\
        LR sched. & \multicolumn{4}{c}{cosine} \\
        Max LR    & \multicolumn{4}{c}{$10^{-3}$}
        \\ \midrule
        ODE solver & \multicolumn{4}{c}{Runge-Kutta 4 (20 steps)} \\     
        \bottomrule
    \end{tabular} \end{small}
    \caption{Parameters for the shape networks in Sec.~\ref{sec:dream_shape}, for the full and the latent space.}
    \label{tab:vit_params}
\end{table}
%--------------------------------------------------------------

%----------------------------------------------------------
\begin{table}[h!]
    \centering
    \begin{small} \begin{tabular}[t]{lc}
    \toprule
    Parameter & Value \\
    \midrule
    Optimizer & Adam \\
    Learning rate & $2\cdot10^{-4}$ \\
    Batch size & 1000 \\
    Epochs & 200 \\
    Number of layers & 3 \\
    Hidden nodes & 512 \\
    Dropout & $20\%$ \\
    Activation function & leaky ReLU \\
    Training samples & 70k \\
    Validation samples & 10k \\
    Testing samples & 20k \\
    \bottomrule
    \end{tabular} \end{small}
    \caption{Parameters for HL and LL classifiers network used to calculate the weights of Fig.~\ref{fig:weights}. The other classifiers use the same hyperparameters but without any dropout.}
    \label{tab:cls_params}
\end{table}
%----------------------------------------------------------
%----------------------------------------------------------
\begin{table}[h!]
    \centering
    \begin{small} \begin{tabular}{lcc} 
    \toprule
        Parameter & \multicolumn{2}{c}{Value} \\ 
        \midrule
         & DS2 & DS3 \\
         \cmidrule{2-3}
        Loss  &  \multicolumn{2}{c}{BCE + $\beta$KL} \\
        $\beta$ & \multicolumn{2}{c}{$10^{-6}$} \\
        Epochs & \multicolumn{2}{c}{200} \\
        Out activation & \multicolumn{2}{c}{sigmoid} \\
        Lr sched. & \multicolumn{2}{c}{OneCycle} \\
        Max lr  & \multicolumn{2}{c}{$10^{-3}$} \\
        \# of blocks &  \multicolumn{2}{c}{2 (+ bottleneck)} \\
        Channels  &  \multicolumn{2}{c}{(64, 64, 2)} \\
        Dim. bottleneck & (2, 15, 9, 9) & (2, 9, 26, 16) \\
        Kernels   & [(3,2,1), (1,1,1)] & [(5,2,3), (1,1,1)] \\
        Strides   & [(3,2,1), (1,1,1)] & [(2,2,1), (1,1,1)]  \\
        Paddings  & [(0,1,0), (0,0,0)] & [(0,1,0), (0,0,0)] \\
        Normalized cut & \multicolumn{2}{c}{$1\cdot10^{-6}$} \\
        \bottomrule
    \end{tabular} \end{small}
    \caption{Parameters of the autoencoder for DS2 and DS3 used for the laViT network in Sec.~\ref{sec:dream_lat}.}
    \label{tab:ae}
\end{table} 
%----------------------------------------------------------

%----------------------------------------------------------
\begin{table}[h]
    \centering
    \begin{small} \begin{tabular}[t]{lc}
    \toprule
    Parameter & Value \\
    \midrule
    Reference solver & midpoint (100 steps) \\
    Initialization & Euler \\
    Optimizer & Adam \\
    Learning rate & $1\cdot10^{-3}$ \\
    Batch size & 100 \\
    Max iterations & 5000 \\
    Stopping patience (iterations) & 200 \\

    \bottomrule
    \end{tabular} \end{small}
    \caption{Parameters used to train BNS solvers, described in Sec.~\ref{sec:dream_bespoke}.}
    \label{tab:bns_params}
\end{table}
%----------------------------------------------------------
\clearpage

%%%%%%%%%%%%%%%%%%%%%%%%%%%%%%%%%%%%%%%%%%%%%%%%%%
\subsection{Autoencoder}
\label{app:ae}

%-------------------------------------------------
\begin{figure}[t]
    \includegraphics[width=0.33\textwidth, page=21]{figs/ae_ds3/ECPhi_layer_dataset_3.pdf} 
    \includegraphics[width=0.33\textwidth, page=1]{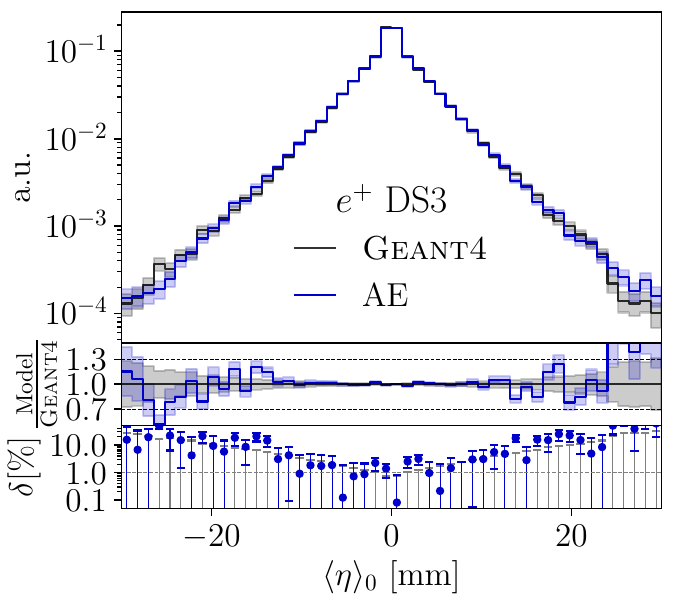} 
    \includegraphics[width=0.33\textwidth, page=11]{figs/ae_ds3/WidthPhi_layer_dataset_3.pdf} \\
    \includegraphics[width=0.33\textwidth, page=3]{figs/ae_ds3/Weighted_Depth_ring_dataset_3_groups_1.pdf}
    \includegraphics[width=0.33\textwidth, page=11]{figs/ae_ds3/Sparsity_layer_dataset_3.pdf} 
    \includegraphics[width=0.33\textwidth, page=1]{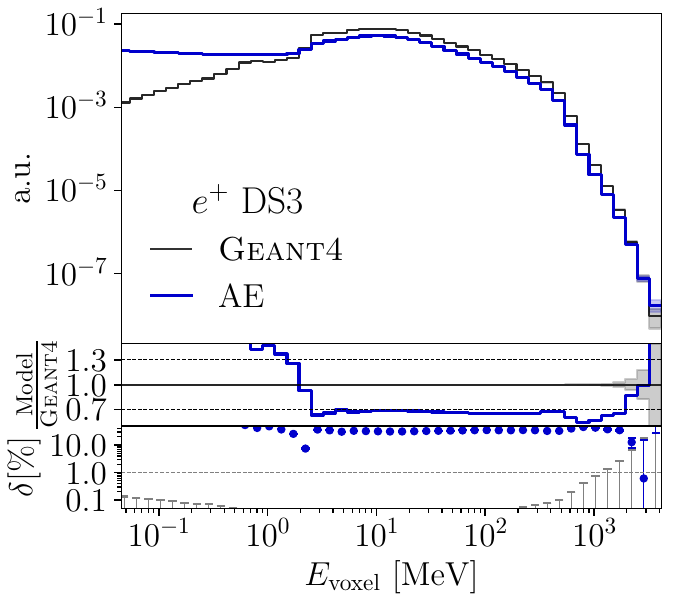} 
    \caption{Selection of high-level features sensitive to the reconstruction of the autoencoder for DS3.}
    \label{fig:ae_hl}
\end{figure}
%-------------------------------------------------

The VAE introduced in  Sec.~\ref{sec:dream_lat} is
trained separately, using the BCE reconstruction loss
\begin{align}
\loss_\text{VAE} = - \Langle x \log (x_\psi) + (1-x)\log (1-x_\psi) \Rangle_{p_\psi(r|x)} + \beta D_\text{KL} [p_\psi (r|x), \normal (0,1)] \;.
\end{align}
This loss provides notably better reconstruction quality than the standard MSE 
loss, both in terms of high-level features and a neural network classifier 
trained to distinguish reconstructed showers from an independent test set.
A detailed description of the network architecture is provided in Tab.~\ref{tab:ae}.
Each block consists of three Conv2d operations that preserve the number of channels of which the final one downsamples according to the stride and padding parameters.
In addition, we break the translation equivariance by adding the coordinates of each input to the activation map as new channels~\cite{Amram:2023onf}.

In Fig.~\ref{fig:ae_hl} we provide a set of kinematic distributions 
similar to Fig.~\ref{fig:ds3_full} for DS3, to illustrate the 
VAE reconstruction. We find that the only missing feature in the learned manifold 
is the distribution of the low-energetic voxels, also reflected in the sparsity. 
We also train a classifier using the hyperparameters of Tab.~\ref{tab:cls_params} on the low-level features which gives an AUC score of 0.512(5) consistently for both, DS2 and DS3.

%%%%%%%%%%%%%%%%%%%%%%%%%%%%%%%%%%%%%%%%%%%%%%%%%%
\subsection{Timing}
\label{app:timing}

%-------------------------------------------------
\begin{table}[]
    \centering
    \begin{small} \begin{tabular}{ccc}
    \toprule
    Network & \multicolumn{2}{c}{Time (ms) on GPU}\\
    \midrule
      & DS2 & DS3\\
    \cmidrule{2-3}
    Energy & 0.37$\pm$0.01 & 0.37$\pm$0.01\\
    Shape (ViT) & 17$\pm$1 & 84$\pm$8 \\
    Shape (LaViT) & 31$\pm$1 & 63$\pm$6\\
    \bottomrule
    \end{tabular} \end{small}
    \caption{Timings for one network forward pass using batch size 100 on an NVIDIA H100.}
    \label{tab:timing}
\end{table}
%-------------------------------------------------
%-------------------------------------------------
%\begin{table}[]
%    \centering
%    \begin{small} \begin{tabular}{ccc}
%    \toprule
%    Network & \multicolumn{2}{c}{Time (ms) on GPU}\\
%    \midrule
%      & DS2 & DS3\\
%    \cmidrule{2-3}
%    Energy & 0.37$\pm$0.01 & 0.37$\pm$0.01\\
%    Shape (ViT) & 0.07$\pm$0.01 & 0.35$\pm$0.01 \\
    %Shape (LaViT) & 31$\pm$1 & 63$\pm$6\\
%    \bottomrule
%    \end{tabular} \end{small}
%    \caption{Timings for one RK4 step using batch size 1 on H100.}
%    \label{tab:timing}
%\end{table}
%-------------------------------------------------
%-------------------------------------------------
%\begin{table}[]
%    \centering
%    \begin{small} \begin{tabular}{ccc}
%    \toprule
%    Network & \multicolumn{2}{c}{Time (s) on CPU}\\
%    \midrule
%      & DS2 & DS3\\
%    \cmidrule{2-3}
%    Energy & 0.02$\pm$0.01 & 0.02$\pm$0.01\\
%    Shape (ViT) & 0.2$\pm$1 & 0.5$\pm$8 \\
    %Shape (LaViT) & 31$\pm$1 & 63$\pm$6\\%
%    \bottomrule
%    \end{tabular} \end{small}
%    \caption{Timings for one RK4 step using batch size 1 on bigmem CPU.}
%    \label{tab:timing}
%\end{table}
%-------------------------------------------------

In Sec.~\ref{sec:sampling}, we study the sampling cost of networks in terms of the number of function evaluations $n_{\text{eval}}$. Here we provide timing measurements for a single forward pass of each of our CFM networks, using a batch size 100. We ran tests using a single NVIDIA H100 GPU and summarize the results in Tab.~\ref{tab:timing}. The times for the energy network are identical across the two datasets since there is no change in the network architecture. Also note that since the energy model is autoregressive, sampling with an $N$-step solver uses $N\times L$ function evaluations, where $L$ is the number of calorimeter layers. As we did not perform an extensive hyperparameter search, we expect there to be room for improvements for all listed models.
%%%%%%%%%%%%%%%%%%%%%%%%%%%%%%%%%%%%%%%%%%%%%%%%%%
\bibliography{refs}
\end{document}